\documentclass[sort&compress]{elsarticle}

\usepackage{lineno,hyperref}
\modulolinenumbers[5]

\usepackage{multirow}
\usepackage{float}

\usepackage{color} %new package

\begin{document}

\begin{frontmatter}
\title{Evaporation of sessile ethanol-water droplets on a highly inclined heated surface}
\author{Pallavi Katre$^{\dagger}$}
\author{Pradeep Gurrala$^{\dagger\dagger}$}
\author{Saravanan Balusamy$^{\dagger\dagger}$}
\author{Sayak Banerjee$^{\dagger\dagger}$}
\author{Kirti Chandra Sahu$^{\dagger\dagger}$\fnref{ksahu@iith.ac.in}}
 \address{
$^{\dagger}$Department of Chemical Engineering, Indian Institute of Technology Hyderabad, Sangareddy 502 285, Telangana, India\\
        $^{\dagger\dagger}$Department of Mechanical and Aerospace Engineering, Indian Institute of Technology Hyderabad, Sangareddy 502 285, Telangana, India}

\begin{abstract}
The evaporation of a binary sessile ethanol-water droplet on an inclined substrate is studied experimentally just below the critical sliding angles for different substrate temperatures. A customized goniometer equipped with a CMOS camera and an infrared (IR) camera is used. The droplet is observed to remain pinned in the advancing side during the evaporation process, while the receding side contracts. The asymmetry in the advancing and receding contact angles of the droplet on inclined substrate results in complex thermo-solutal Marangoni convection that is captured through IR images. The droplet exhibits two distinct oscillatory water-rich cold regions around the advancing contact line during the early stage of evaporation, while the more volatile and lighter ethanol creates a hotter and rapidly evaporating cell near the receding side. As ethanol evaporates away, the ethanol rich cells collapse producing thermal pulsations along the incline. Subsequently, the thermal patterns become similar to that of the pure-water droplet. It is also observed that the thermo-solutal driven oscillatory convection increases with increasing substrate temperature. Despite the complexity in convection dynamics, the evaporation rate exhibits a universal behavior in the normalized time at different substrate temperatures which can be represented by piecewise linear fits at the early and late stages of evaporation. 
\end{abstract}


\end{frontmatter}

\section{Introduction} \label{sec:intro}

The study of droplet evaporation has received considerable attention in the research literature due to its importance in many industrial applications, such as inkjet printing\cite{lim2012deposit,kim2006direct, park2006control,lim2009experimental,koo2006fabrication,tekin2004ink,de2004inkjet}, hot spot cooling, droplet-based microfluidics\cite{deegan1997capillary}, coating technology\cite{yanagisawa2014investigation}, and even natural phenomena\cite{tripathi2015evaporating}. The wetting characteristics of sessile and pendant drops have also attracted researchers due to their abundant fundamental and industrial relevance (see e.g. Refs.\cite{langmuir1918evaporation,picknett1977evaporation}). Several researchers have studied the evaporation of sessile droplets of pure liquids on flat surfaces\cite{sobac2012thermal,birdi1989study,sobac2012thermocapillary}, which has recently been reviewed in Brutin and Starov\cite{brutin2018recent}.

In addition, few researchers have investigated the dynamics of sessile droplets of pure fluids on inclined surfaces, as they exhibit unequal advancing and receding contact angles due to the effects of gravity\cite{extrand1995liquid,kim2002sliding,chou2012drops,kim2017evaporation,annapragada2012droplet,yilbas2017dynamics,ellaban2017instability}. Extrand and Kumagaya \cite{extrand1995liquid} experimentally investigated droplet shapes at different tilt angles of the substrate. For a stationary droplet (i.e., below the critical angle at which it starts to migrate due to gravity), they investigated the dependence of the contact angle hysteresis (i.e., the difference between the advancing and receding contact angles) on droplet shape and found a linear relationship between them. Chou {\it et al.} \cite{chou2012drops} conducted experiments and numerical simulations to study the wetting behavior of a droplet on inclined surfaces. The pinning and de-pinning behaviors of the advancing and receding contact angles were observed at various inclination angles. It was found that increasing the tilt angle decreases the surface free energy. Annapragada {\it et al.} \cite{annapragada2012droplet}, through experiments and numerical simulations, investigated the shape of a sessile droplet on a hydrophobic surface at various inclination angles. Few researchers \cite{kim2002sliding,yilbas2017dynamics} have also studied droplet migration on inclined substrates, but this is not the subject of the present study.

All of the above-mentioned studies involving sessile pure droplets on inclined substrates have focused on shape and contact angle hysteresis of the droplets. Kim {\it et al.} \cite{kim2017evaporation} investigated the evaporation of deionized water droplets on a tilted glass surface for different tilt angles. They observed that droplet deformation due to gravity increased the difference between the advancing and receding contact angles, which in turn increased the evaporation time.

Next, we discuss the evaporation dynamics of sessile droplets of binary mixtures on flat and inclined substrates. Sefiane {\it et al.} \cite{sefiane2003experimental} experimentally investigated the evaporation dynamics of water-ethanol binary droplets of different compositions on a flat surface at ambient temperature. The evaporation dynamics of binary droplets was compared with that of pure droplets of each component. Unlike pure water or pure ethanol droplets, it is observed that in the case of binary droplets, the evaporation occurs at three different stages. The more volatile ethanol evaporates quickly in the first stage, but less volatile water evaporates relatively slowly in the last stage. In the intermediate stage, the droplet volume remains almost constant as the wetted radius and the contact angle change simultaneously. Subsequently, these distinct evaporation steps were explained in the literature\cite{christy2011flow,bennacer2014vortices} by investigating the flow field using particle image velocimetry (PIV). Some researchers have also investigated the effect of surface properties on the evaporation of various binary mixtures \cite{ozturk2018evaporation,schofield2018lifetimes,feng2017octagon,shi2009wetting,sefiane2008wetting,wang2008evaporation,cheng2006evaporation}. The influence of Marangoni flow on the wetting dynamics and the evaporation of binary sessile droplets on flat surfaces have also been studied in the past\cite{karpitschka2017marangoni,karapetsas2016evaporation,karapetsas2014thermocapillary}. Recently, Gurrala {\it et al.} \cite{gurrala2019evaporation} studied the effect of substrate temperature and composition on the evaporation of ethanol-water binary sessile droplets on a flat substrate. They observed an early spreading stage, an intermediate pinning stage, and a late receding stage of evaporation at elevated substrate temperatures, leading to non-monotonic evaporation trends at high ethanol concentrations. They also performed theoretical modeling that accounts for diffusion, free convection, and passive transport due to free convection of air, that satisfactorily predicts the evaporation rates observed in their experiments. Yonemoto {\it et al} \cite{yonemoto2018sliding} studied the sliding behavior of water-ethanol droplets on an inclined surface with low surface energy. The critical angle of inclination ($\alpha$) of the substrate at which the droplet begins to slip is obtained as a function of the surface energy density, and the adhesion that determines the onset of the advancing and receding contact angles has been observed to exhibit a linear relationship with the surface energy density of the liquid.

The evaporation of binary droplets is complex due to the differences in the volatility of the components in the mixture, and the associated Marangoni convection and hydrothermal waves. Thus, few researchers have used infrared (IR) thermography to study the thermal pattern of water-ethanol sessile droplets at room\cite{innocenzi2008evaporation} and elevated substrate temperatures\cite{saenz2017dynamics}. S{\'a}enz {\it et al.} \cite{saenz2017dynamics} considered the evaporation of non-spherical droplets on flat substrates. Using a similar setup, Mamalis {\it et al.} \cite {mamalis2016motion} studied the temporal evolution of the thermal pattern of the free surface of a sessile water + 1 - butanol (5\% by volume) droplet placed on a 5$^\circ$ inclined substrate maintained at 60$^\circ$C. The substrate is overlaid on a silicone oil layer and the binary droplet is allowed to slide on the oil layer. The binary mixture considered exhibits a non-monotonic surface tension - temperature dependency. Therefore, in some cases, they observed an upward movement of the droplet due to the thermocapillary forces acting against gravitational force. It was found that the interaction between thermocapillary and body forces also increases the spreading rate of the binary droplet.

The above review indicates that while many researchers have studied the dynamics of evaporation of binary droplets at room temperature on flat substrates, little attention has been paid to inclined substrates at elevated temperatures. To the best of our knowledge, the research work of Mamalis {\it et al.} \cite{mamalis2016motion} was the only study in this direction but was performed at a low angle of inclination (5$^\circ$). In contrast, the present work focuses on the evaporation dynamics of a water/ethanol binary droplet on a substrate inclined just below the critical sliding angle at different substrate temperatures using optical and IR imaging techniques. The composition of water (W) - ethanol (E) binary mixture is fixed at E 20\% + W 80\% and the initial volume of the droplet is 5 $\mu$l in all the experiments performed in the present study. It is found that the critical angle of the inclination ($\alpha$) is a function of the substrate temperature ($T_s$). The evaporation dynamics of the binary droplet and associated thermal patterns at various elevated temperatures are analyzed. We have also developed a semi-empirical correlation between the evaporation rate and the normalized lifetime of the binary droplet at different substrate temperatures. 

The rest of the paper is organized as follows. The experimental set-up and procedure are discussed in Section 2. The results obtained from our experiments are presented in Section 3, and concluding remarks are given in Section 4.

\section{Experimental set-up} \label{sec:expt}

\begin{figure}[h]
\hspace{4cm} (a)  \hspace{5.0cm} (b) \\
\begin{minipage}{0.7\textwidth}
\hspace{0.1cm} \includegraphics[width=0.9\textwidth]{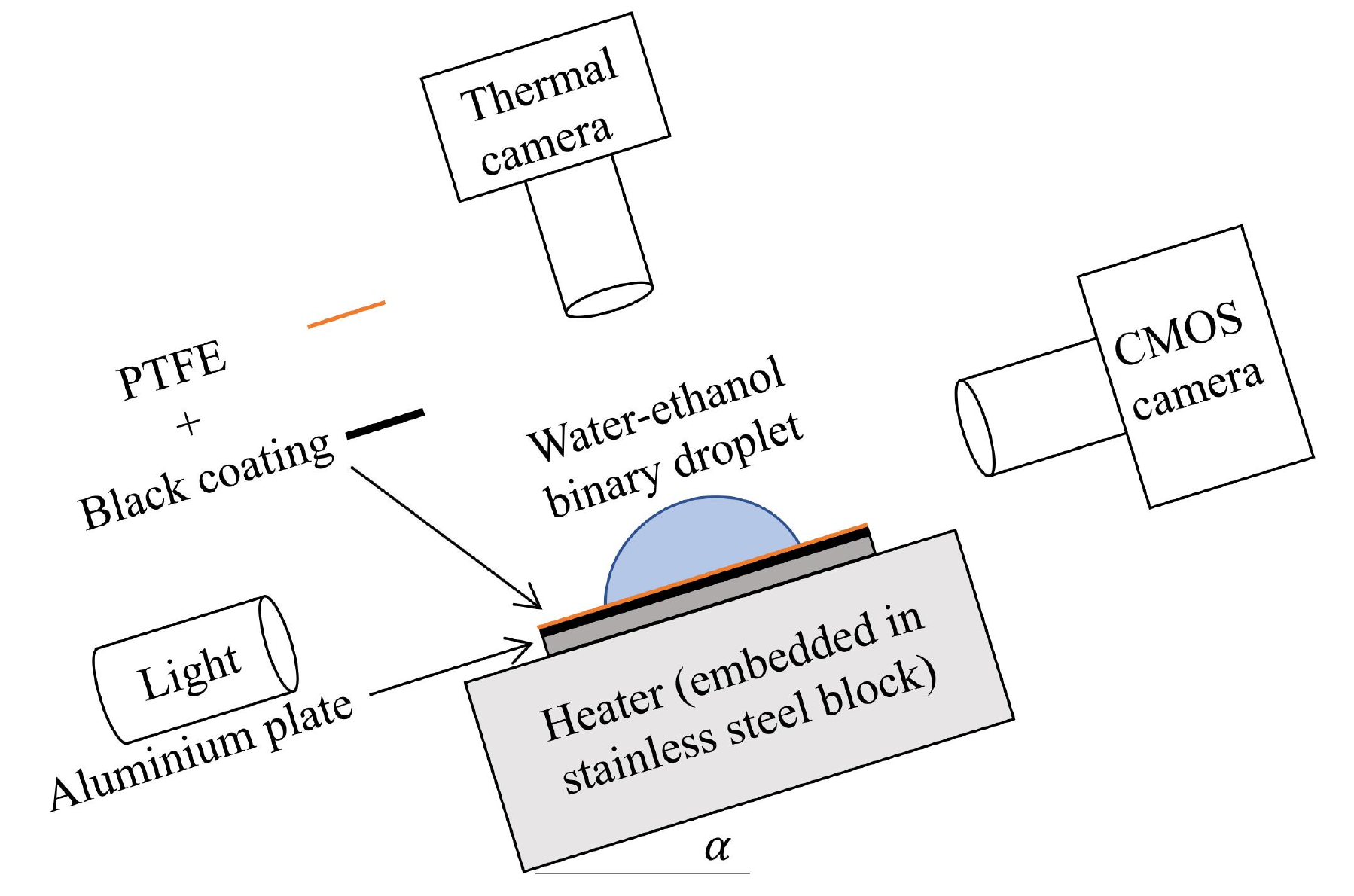}
\end{minipage}\hfill
\begin{minipage}{0.3\textwidth}                
\resizebox{0.9\textwidth}{!}{
\begin{tabular}{|c|c|} 	\hline
$T_s$ ($^{\circ}$C)  &  $\alpha$ (degree)  \\ \hline
40   &  37    	\\ \hline
50   &  40  	\\ \hline
60   &  43   	\\ \hline
70   &  45      \\ \hline
\end{tabular}
}
\end{minipage}
\caption{(a) Schematic of the experimental set-up that consists of a heater, a Polytetrafluoroethylene (PTFE) substrate placed on a stainless steel plate, a light source, IR and CMOS cameras. (b) The critical angle of inclination ($\alpha$, in degree) of a 5 $\mu$l binary droplet at the onset of sliding at different substrate temperatures, $T_s$.}
\label{fig:geom}
\end{figure}

The evaporation dynamics of an ethanol (E) -water (W) binary droplet of composition (E 20\% + W 80\%) by volume is investigated using a customized goniometer (Make: Holmarc Opto-Mechatronics Pvt. Ltd.). For each substrate temperature, the inclination of the substrate $(\alpha)$ is kept just below the critical inclination angle at which the droplet begins to migrate downward due to gravity. The goniometer setup consists of a multilayer metal block, a motorized pump to generate the droplets, and a proportional-integral-derivative (PID) controller that maintains the substrate temperature. It is also equipped with a light source and a complementary metal-oxide-semiconductor (CMOS) camera (Make: Do3Think, Model: DS-CBY501E-H) to record the side view of the droplet, and an infrared (IR) camera (Make: FLIR, Model: X6540sc) that captures the top view of the droplet in the direction perpendicular to the substrate at each inclination angle. A schematic diagram of the experimental setup is shown in Fig. \ref{fig:geom}(a). The entire setup is placed inside the goniometer box to minimize external disturbances. The goniometer is also equipped with a rotation mechanism to set the inclination angle of the substrate. The experiments are performed at a constant ambient temperature of $21^\circ$C in an air-conditioned room. The relative humidity measured using the hygrometer (Make: HTC Instruments, Model: 288-ATH) was found to be $70 \pm 10$ \% throughout the experiments.\\

The multilayer metal block consists of a base stainless steel metal block embedded with two electric heaters controlled by a PID controller, and an aluminum plate coated with black paint of size 100mm $\times$ 80mm $\times$ 15 mm. Polytetrafluoroethylene (PTFE) tape of thickness of 100 $\mu$m is used as a substrate that is placed on an aluminum plate. We use black paint between the PTFE substrate and aluminum plate to minimize the reflection in IR images. The stability of the PTFE substrate at different temperatures is checked using an optical microscope imaging system after heating the substrate for 10 minutes. {In Fig. 14 (supporting information)}, it can be seen that the highest roughness value is less than 9 $\mu$m even at $T_s = 70^\circ$C. This indicates that the PTFE substrate is stable and exhibits uniform roughness at all temperatures considered in this study. A `K- type' thermocouple (Make: OMEGA Engineering Singapore) is used to verify the temperature of the substrate. A heat-flux meter (Make: Captec Entreprise, France) having a sensitivity of 1.17 $\mu{\rm V m}^2$/W and a diameter of 10 mm is used to measure the heat-flux through the droplet.

In each experiment, the PTFE substrate is cleaned with isopropanol, blow-dried using an air compressor, and then applied to the aluminum plate. The experiment is then performed by placing a sessile binary droplet on the substrate after it attains a steady-state at each high-temperature condition. The homogeneous ethanol-water binary solution (E 20\% + W 80\%) is prepared on a volume basis using absolute ethanol (99.9\% purity) and deionized water (purity 8:2 M$\Omega$) with the help of a stirrer. A binary droplet of the initial volume of 5 $\mu$l is created using a 250 $\mu$l U-tek (Unitek Scientific Corporation) chromatography syringe with a 2.47 mm diameter needle attached to a motorized pump that controls the droplet flow rate and volume with a $\pm$1\% error. The experiments are performed at four values of substrate temperature, $T_s = 40^\circ$C, $50^\circ$C, $60^\circ$C and $70^\circ$C. The critical angle of the inclination is calculated for all the substrate temperatures by conducting experiments starting with an inclination angle of 10$^\circ$ and incrementing it 5$^\circ$ until the droplet starts to slide down at a specific angle. To get the exact critical inclination angle, an increment of 1$^\circ$ was used near the critical angle. Fig. \ref{fig:geom} (b) shows a list of substrate temperatures and the corresponding critical angles at which the rest of the experiments are performed. It can be seen that the critical angle of inclination increases with increasing the substrate temperature.

The temporal evolution of the side view of droplet evaporation is acquired using the CMOS camera with a spatial resolution of $1280 \times 960$ pixels at a frame-rate of 10 frames/second (fps). The instant at which the droplet touches the substrate is considered to be the start of the experiment (i.e., time, $t=0$). The droplet image sequence obtained from the CMOS camera is filtered to remove the random noises, and the droplet outline is then traced using an in-house Matlab code. The image post-processing procedure is the same as that of Gurrala {\it et al.} \cite{gurrala2019evaporation}. 

%2
\begin{figure}
\centering
\includegraphics[width=0.4\textwidth]{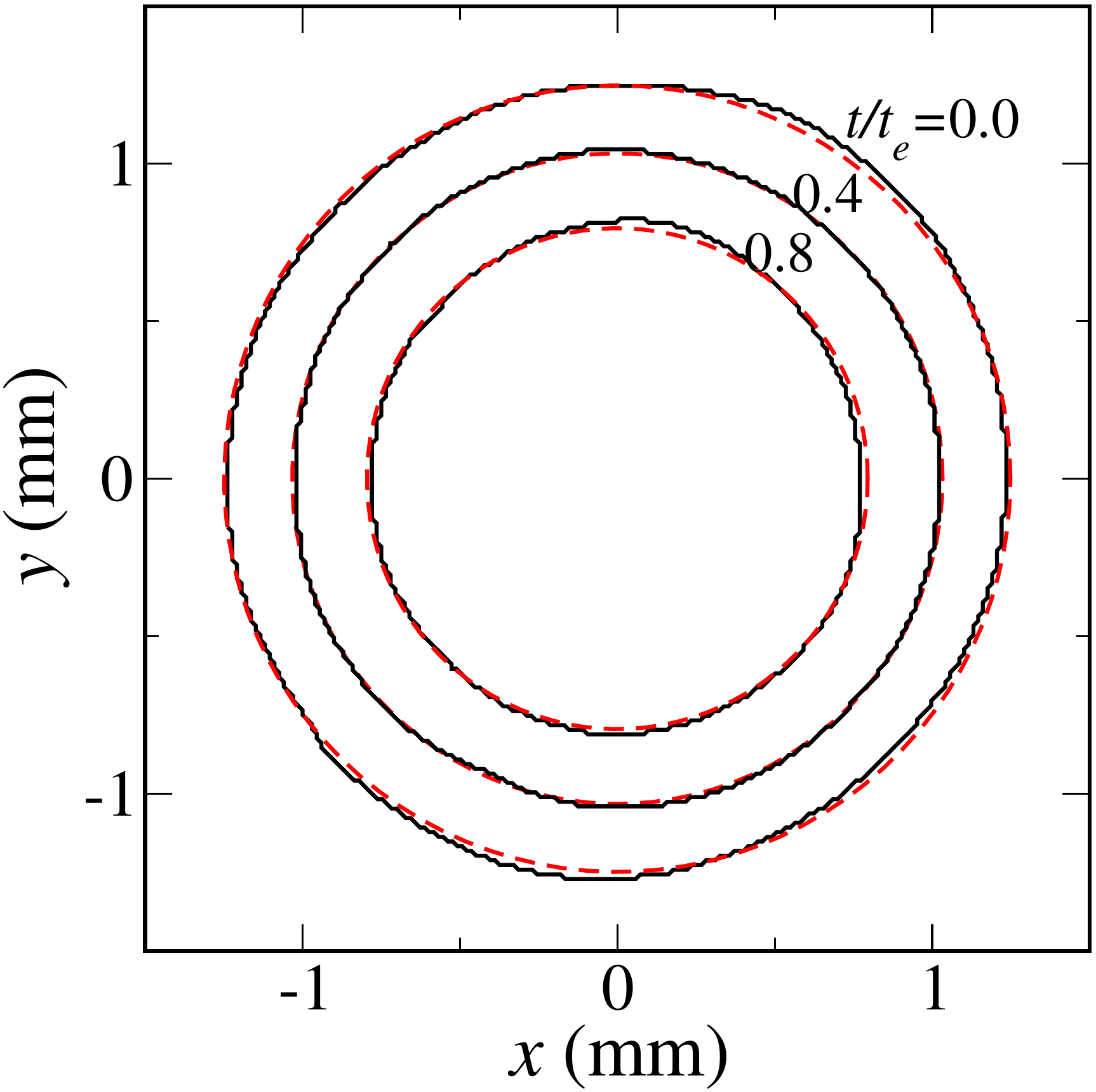}
\caption{Demonstration of the circular nature of the top view of the binary (E 20\% + W 80\%) droplet (solid black line) at $T_s=50^\circ$ at different values of the dimensionless time, $t/t_e$. The fitted circle at each value of $t/t_e$ is shown by red dotted line.}
\label{fig:fig2}
\end{figure}

\begin{figure}
\centering
\includegraphics[width=0.5\textwidth]{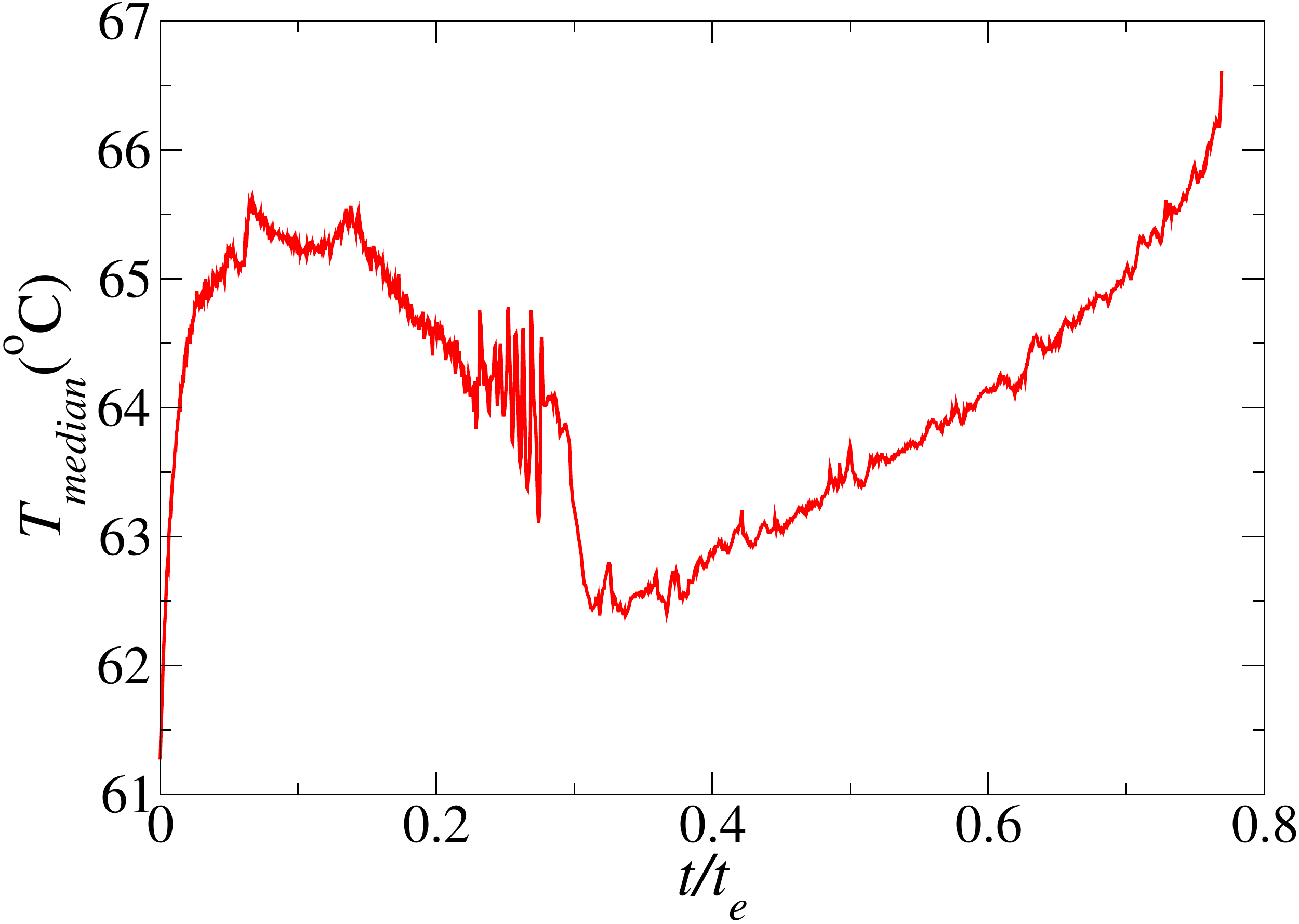}
\caption{Temporal variation of median value of the temperature of a $5 \mu$L binary (E 20\% + W 80\%) droplet at $T_s = 70^\circ$C.}
\label{fig:fig3}
\end{figure}

An infrared camera captured the thermal patterns at the liquid-air interface of the droplet as it evaporates. The IR camera acquired images in the spectral range of 3 $\mu$m - 5 $\mu$m with a resolution of $640 \times 512$ pixels. The spatial resolution is 67.11 pixel/mm. The lens used is the macroscopic G1 WD30 with F/3.0 aperture and 26 cm focal length. The recording of the thermal images is performed at a frame rate of 50 Hz. A specialized software (FLIR ResearchIR Max, Version 4.40.9.30) is used to convert the image to a temperature field using the emissivity parameters embedded in the software. The emissivity of the ethanol-water binary mixture is close to the emissivity of pure water \cite{mamalis2016motion}, and thus an emissivity value of 0.94 is used in the present study. The droplet temperature field obtained from the IR image is then post-processed using an in-house Matlab code. Fig. \ref{fig:fig2} shows the wetted perimeter of the droplet at different evaporation stages at $T_s = 50^\circ$C and the corresponding fitted circles that indicate the circularity of the perimeter even at 80\% of the lifetime $(t_e)$ of the droplet. A similar trend is observed at all elevated substrate temperatures considered in our study. The evolution of the median temperature of the binary droplet at a substrate temperature of $70^\circ$C is shown in Fig. \ref{fig:fig3}. It can be seen that following the rapid heating of the droplet, a slight decrease in temperature is observed due to the thermo-solutal Marangoni convection, which brings the cold water to the surface.  At about 20\% of the lifetime of the droplet, instability in the thermal pattern is observed, which facilitates the rapid mixing of the hot and cold regions, followed by a period of slow heating of the droplet as the temperature gradient is reduced due to the evaporation.

%3

The experiments to measure the heat flux entering the droplet were performed separately from the experiments used to calculate the droplet height, radius, and contact angles. A heat flux sensor (Make: Captec Entreprise, France) with a diameter of 10 mm and a sensitivity of 1.17 $\mu$Vm$^2$/W is used. The sensor is coated with a black paint to avoid the variation in the emissivity. The sensor is connected to a data acquisition system (National Instruments cDAQ-9178 with a NI 9213 input module) to record the measurements. A PTFE sheet is then placed on the sensor and the experiments are carried out similarly like other experiments as mentioned above. The convective heat flux data is measured for all substrate temperatures, which are discussed later in the next section. 

For each test condition, the temporal variations of the wetted diameter $(D)$, height $(h)$, advancing contact angle ($\theta_a$) and receding contact angle ($\theta_r$) are extracted from the side view of the droplet. At the beginning of each new experiment, a new PTFE substrate is used and the syringe is cleaned with acetone and then dried. Each experiment is repeated at least four times at all substrate temperatures to ensure reproducibility. 

\section{Results and discussion}
\label{sec:dis}

%\subsection{3.1. Droplet lifetime}

\begin{figure}
\centering
\includegraphics[width=0.5\textwidth]{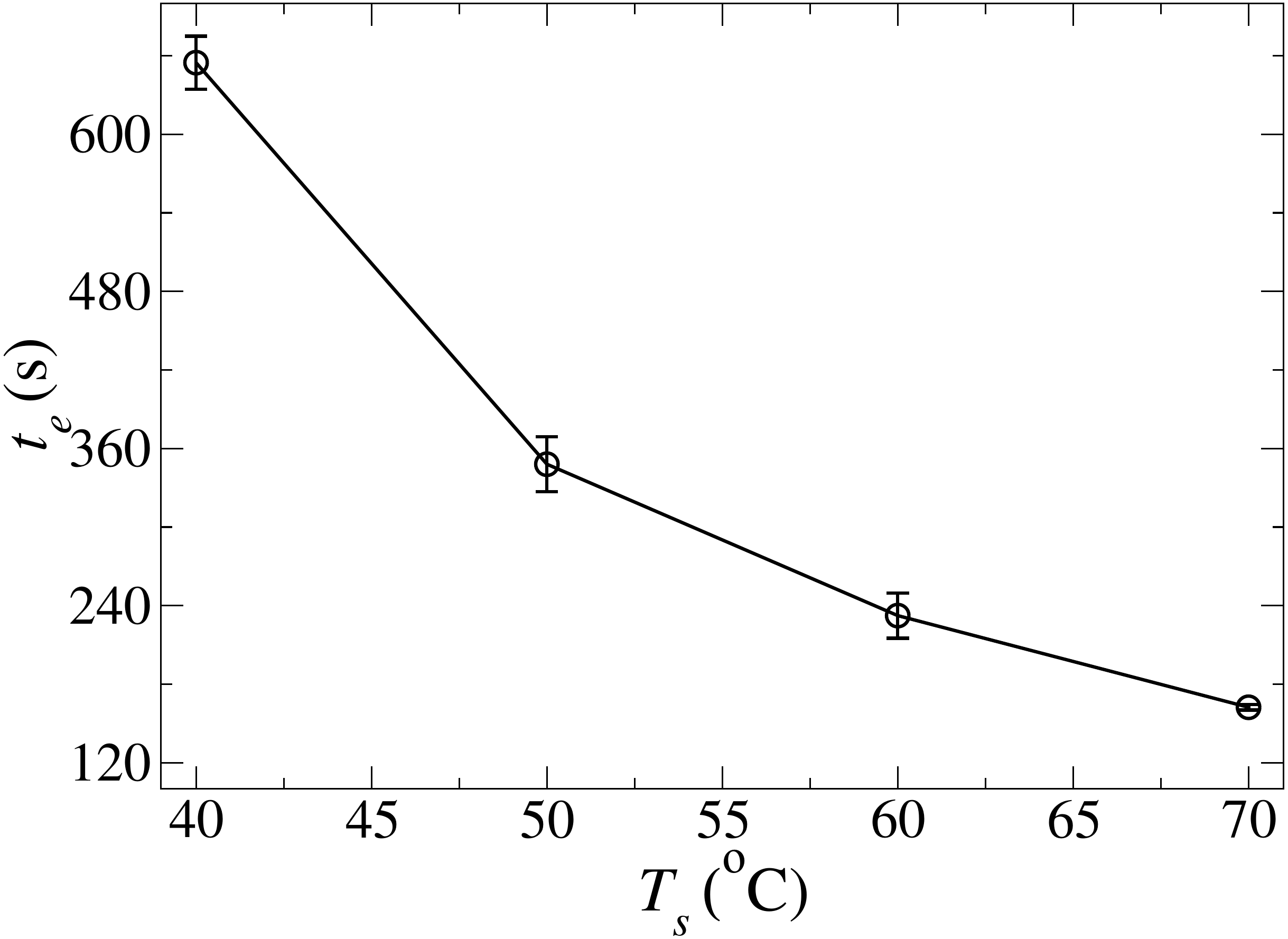}
\caption{Variation of the lifetime $(t_e)$ of a binary (E 20\% + W 80\%) droplet with the substrate temperature, $T_s$. The inclination of the substrate is varied with the substrate temperature, $T_s$ as given in Fig. \ref{fig:geom}(b).}
\label{fig:fig2a}
\end{figure}

We begin by presenting the evaporation dynamics of a binary (E20 \% + W80 \%) sessile droplet of initial volume 5$\mu$l placed on an inclined substrate at the critical angles for different substrate temperatures $(T_s)$. The lifetime of the droplet $(t_e)$, which is defined as the duration between the instants when the droplet touches the substrate and when it evaporates completely from the substrate, is measured. Fig. \ref{fig:fig2a} shows the lifetime of the binary droplet at different substrate temperatures. As expected, increasing the substrate temperature accelerates the evaporation process, thereby shortening the lifetime of the binary droplet. It has been observed that the lifetime of a binary droplet on an inclined surface is shorter than that observed on a flat substrate $(\alpha=0)$, where other conditions are almost identical. For example, at $T_s = 50^\circ$C, in the case of an inclined substrate $(\alpha=40^\circ)$ the lifetime of a binary (E20 \% + W80 \%) sessile droplet, $t_e = 348\pm20.9$ seconds, but on a flat substrate $(\alpha=0^\circ)$ $t_e=458.3\pm15.9$ seconds.

%4

%5
\begin{figure}
\centering
\hspace{0.6cm}  (a) \hspace{5.8cm} (b) \\
\vspace{-0.8cm}
\includegraphics[width=0.48\textwidth]{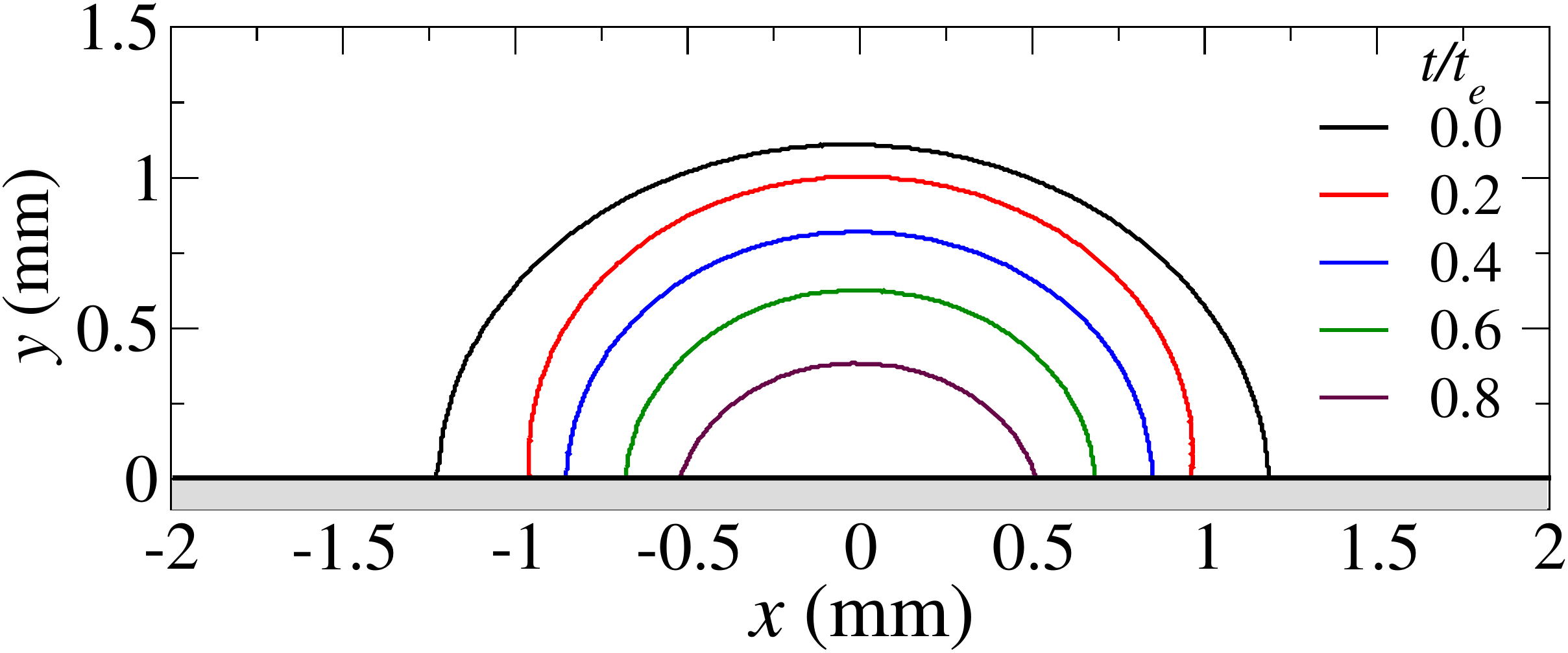} \hspace{2mm} \includegraphics[width=0.48\textwidth]{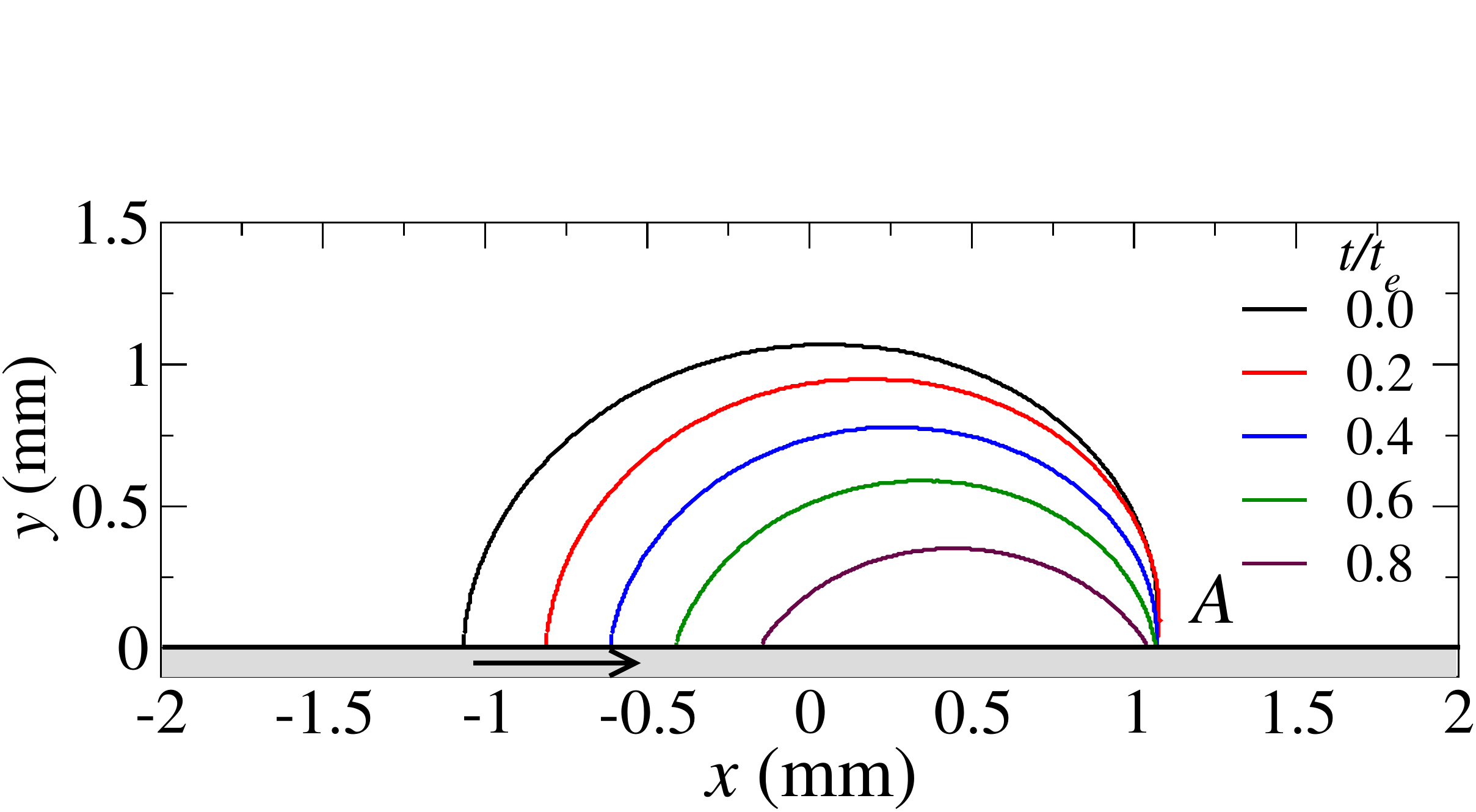}\\
\caption{Evolution of binary (E 20\% + W 80\%) droplet profile on (a) flat and (b) inclined substrates at $T_s=50^\circ$C. The arrow line in panel (b) indicates the direction in which the droplet moves, and the marked ``A''  denotes the advancing side of the droplet.}
\label{fig:fig5}
\end{figure}

%\subsection{3.2. Flat versus inclined substrate}

\begin{figure}
\centering
\includegraphics[width=0.5\textwidth]{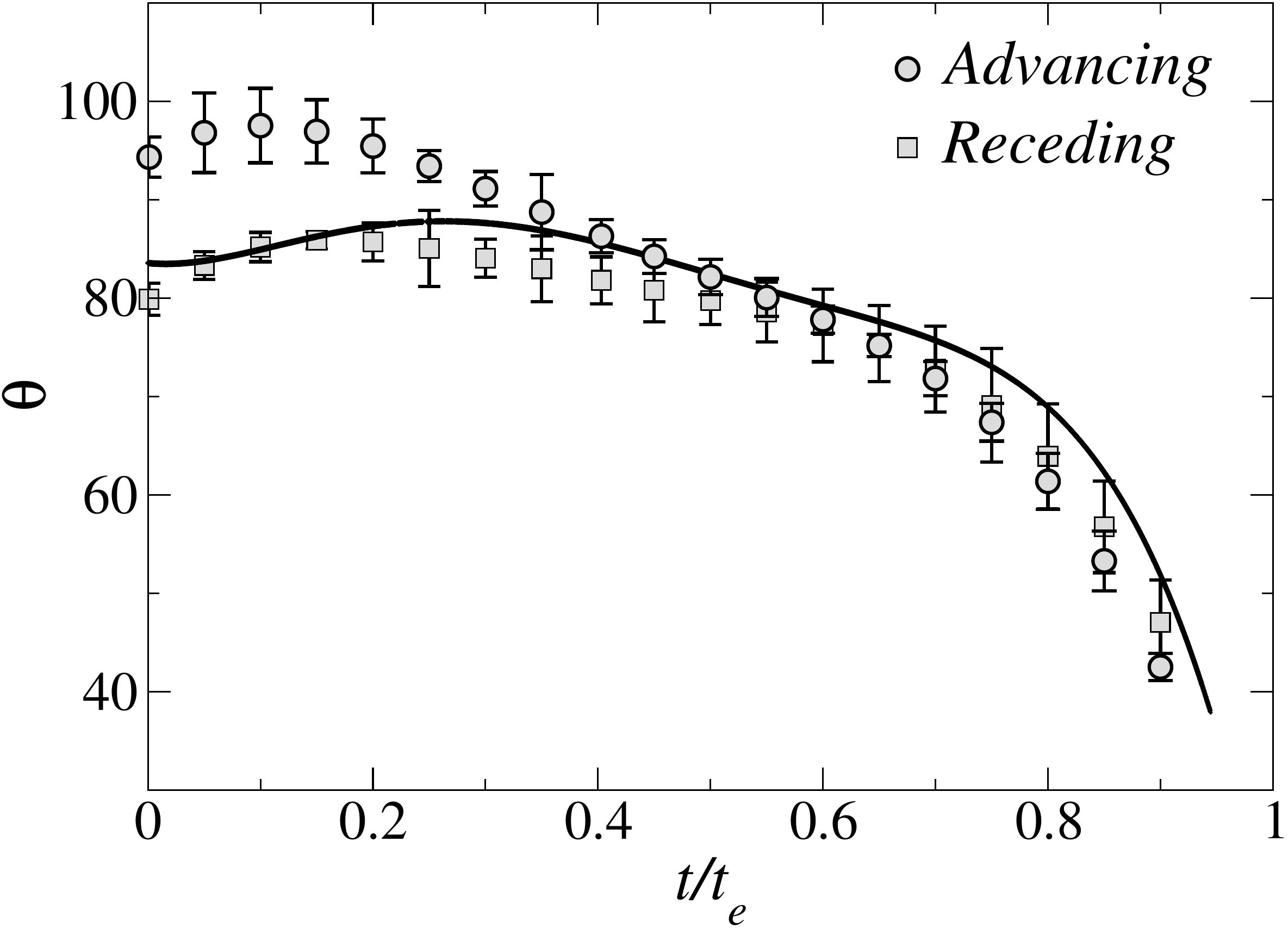}
\caption{Evolution of the advancing and receding contact angles of binary (E 20\%  + W 80\%) droplets on an inclined substrate at $T_s =50^\circ$C. The continuous line represents the contact angle of a droplet on a flat substrate under the same conditions.}
\label{fig:fig-S2}
\end{figure}

In order to contrast the evaporation dynamics of a droplet on a flat $(\alpha=0)$ and an inclined ($\alpha=40^\circ$) substrates, the profiles of the binary (E 20\% + W 80\%) droplet at different normalised times, $t/t_e$ are shown in Figs. \ref{fig:fig5}(a) and (b), respectively. In both the cases, $T_s= 50^\circ$C. In the case of a flat substrate (Fig. \ref{fig:fig5}(a)), it can be seen that the droplet remains axisymmetric as it evaporates indicating that evaporation near the contact line is uniform over the lifetime. On the other hand, in the case of the inclined substrate (Fig. \ref{fig:fig5}(b)), the droplet exhibits contact angle hysteresis due to the effect of gravity. It can be seen that at the early stage of evaporation, the advancing contact angle is greater than the receding contact angle of the droplet, as shown in Fig. \ref{fig:fig-S2}. As time progresses, as the droplet size decreases, the receding side contact line moves down the inclination slope, while the advancing contact angle remains pinned.  This indicates that the evaporation is more in this receding side as compared to the advancing side. At the later stage of evaporation $(t/t_e>0.6)$, the contact angle hysteresis disappears (Fig. \ref{fig:fig-S2}) and the droplet becomes axisymmetric (Fig. \ref{fig:fig5} (b)).

%6

%\subsection{3.3. Evaporation of Pure Water Versus Binary (E 20\% + W 80\%) Droplet}
%7
\begin{figure}
	\centering
	(a) \\
	\includegraphics[width=0.98\textwidth]{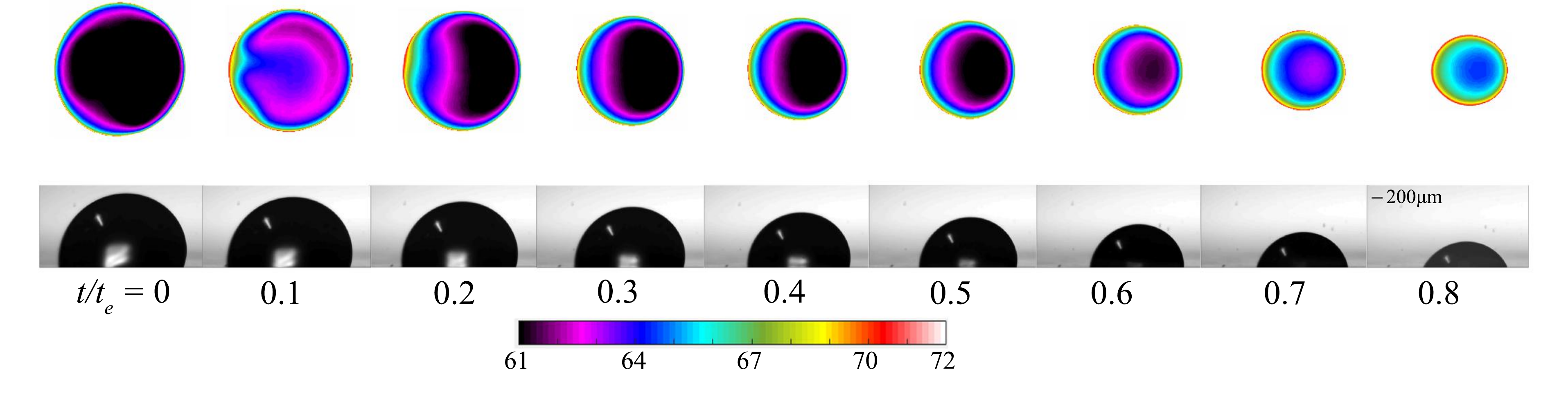}
	(b) \\
	\includegraphics[width=0.98\textwidth]{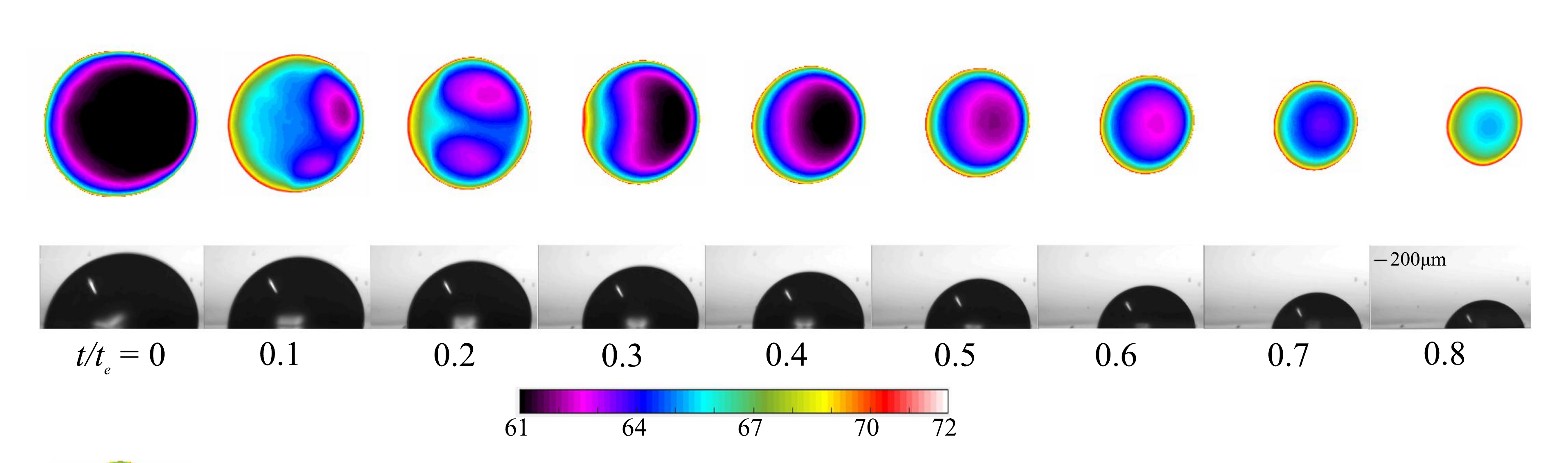} 
	\caption{Temporal evolutions of the evaporation dynamics of (a) pure water (E 0\% + W 100\%), and (b) binary (E 20\% + W 80\%)  droplets at $T_s=70^\circ$C and $\alpha = 45^{\circ}$. The top row in each panel shows the temperature pattern at the liquid-air interface of the droplet captured by the IR camera. The bottom row presents the side view of the droplet captured by the CMOS camera.}
	\label{fig:fig7}
\end{figure}

Next, we examine the evaporation dynamics of pure water and binary (E 20 \% + W 80 \%) droplets at $T_s=70^\circ$C with the corresponding critical angle of inclination ($\alpha = 45^{\circ}$; see Fig. \ref{fig:geom}b). Apart from the difference in composition, it is to be noted that $45^\circ$ is the critical angle only for the (E 20\% + W 80\%) droplet and not for the pure water droplet, which does not slide even for $\alpha= 90^\circ$. Figs. \ref{fig:fig7}(a) and (b) show the temporal evolution of pure water and binary droplets at an elevated substrate temperature ($T_s=70^\circ$C). In each panel of Fig. \ref{fig:fig7}, the top and bottom rows present the infrared images obtained using the IR camera and the side views taken using the CMOS camera, respectively. Examination of the infrared images at different values of $t/t_e$ reveals the following: (i) The receding contact line side of the droplet is hotter than the advancing contact line side for both pure water and binary droplets. (ii) In the case of a pure water droplet (Figs. \ref{fig:fig7}(a)), the droplet undergoes thermocapillary-driven convection and experiences a non-monotonic temperature gradient in the radial direction during the early stage of evaporation ($t/t_e<0.2$). The droplet then acquires a steady temperature gradient along with the inclination of the substrate. (iii) In contrast to the pure water droplet, the binary droplet shows two distinct cold regions of different sizes during the early stage of evaporation ($t/t_e<0.3$). These regions oscillate around the advancing contact line area {(see Fig. 15; supporting information)}. This may be due to the competition between the evaporation rates of the ethanol and water components of the binary mixture, along with the thermo-solutal Marangoni convection currents, which create zones of ethanol- and water-rich liquid within an initially homogeneous droplet. 

%8
\begin{figure}
\centering
\includegraphics[width=0.75\textwidth]{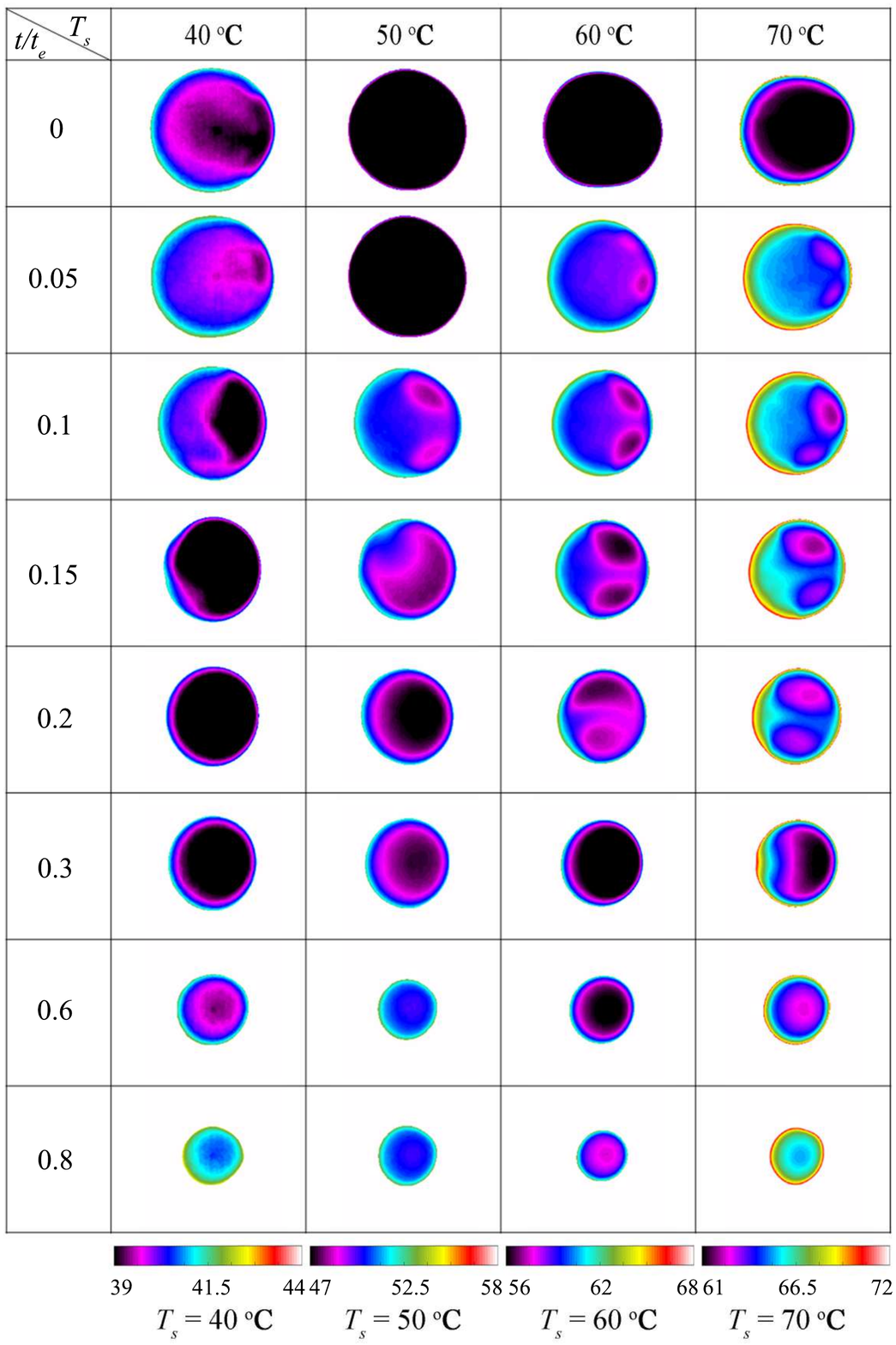}
\caption{Temporal evolutions of temperature contours of the binary (E 20\% + W 80\%) droplet at the critical angles at different substrate temperatures. The color-bars indicate the variations in the temperature field (in $^\circ$C).}
\label{fig:IR_Ts}
\end{figure}

As ethanol is more volatile than water, it evaporates faster as it migrates toward the triple contact line, resulting in water-rich globules, which can be seen as two different cold regions that oscillate about the advancing contact line area. These oscillations occur three to four times, causing intense mixing and merging into a single zone as the concentration of ethanol continues to decrease within the binary droplet. At this stage, the remaining ethanol accumulates in the receding contact line area as a hot and rapidly evaporating cell, which eventually collapses due to the thermal pulsation observed around $t/t_e=0.3$ {(Fig. 16; supporting information)}. After this, the thermal pattern of the binary droplet is similar to that of the pure water droplet, indicating that most of the ethanol component has evaporated, leaving only the water component in the binary droplet at a later stage. It is also evident that the surface temperature of the binary droplet is always higher than the pure water droplet because of the stronger convection flow generated during the initial evaporation stage of the binary droplet due to the solutal Marangoni effect. We propose that the asymmetric segregation of ethanol-rich zone in the receding side and the water-rich zone in the advancing front, coupled with the complex convection dynamics observed from thermal imaging, is responsible for the faster evaporation of the binary droplet on an inclined plate as opposed to that of a flat plate. 

%\subsection{3.4. Effect of Substrate Temperature on the Evaporation Rate of Binary (E 20\% + W 80\%) Droplets}
%9
\begin{figure}
	\centering
	\includegraphics[width=0.65\textwidth]{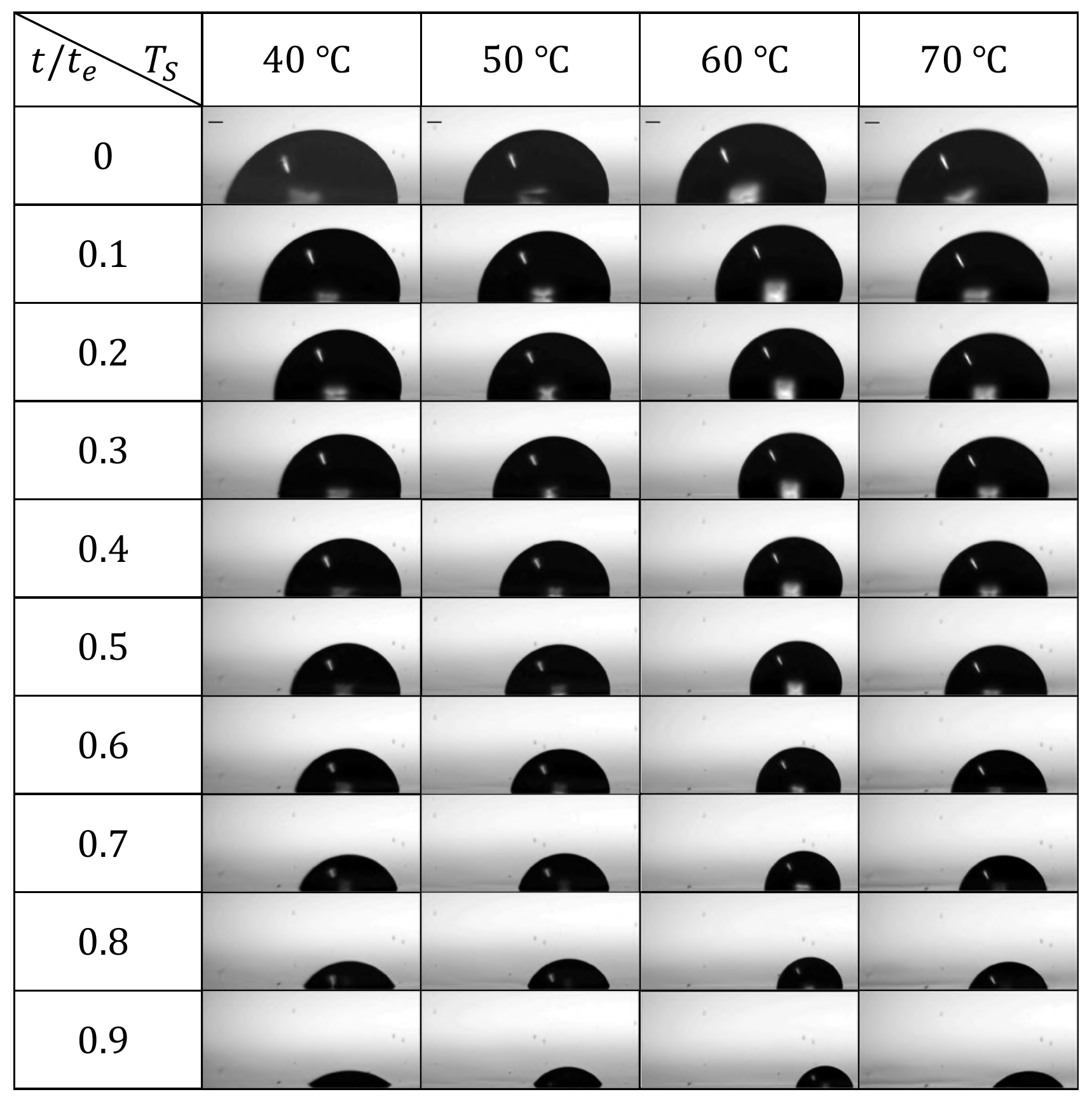} 
	\caption{Temporal evolutions of the side-view of the (E 20\% + W 80\%) droplet at different substrate temperatures, $T_s$ and at the corresponding values of the critical inclination angle ($\alpha$).}
	\label{fig:Side_Ts}
\end{figure}

Then, we study the effect of the substrate temperature, $T_s$ on the evaporation dynamics of a binary (E 20\% + W 80\%) droplet. Fig. \ref{fig:IR_Ts} shows the temporal evolutions of the temperature contour of the binary (E 20\% + W 80\%) droplet at the critical inclination angles of the substrate at various substrate temperatures ($T_s$). In all the cases, at $t/t_e=0$, the temperature at the center of the droplet is lower than that of the triple contact line. As time progresses, the temperature at the center of the droplet increases, but is subjected to convective fluctuations. We observe that the thermo-solutal driven convection intensity increases with the increase in the substrate temperature. For all values of $T_s$, the evaporation is more on the receding side of the droplet. This is because water, which is heavier and less volatile than ethanol, moves to the advancing side due to gravity and the thermo-solutal Marangoni force, causing the ethanol component to move to the receding side of the droplet. Therefore, the ethanol concentration increases on the receding side and evaporates faster due to its higher volatility as compared to water. At $T_s=40^\circ$C, the colder region, that covers the advancing side of the droplet at $t/t_e=0.1$, quickly spreads over the entire top surface of the droplet. For $T_s \ge 50^\circ$C, two distinct oscillating cold regions are evident, and the duration of the oscillatory motion of these regions increases with increasing the value of $T_s$. At $T_s=70^\circ$C, the droplet undergoes extensive thermocapillary convection and exhibits a pulsation phenomenon at $t/t_e \approx 0.3$ (as also discussed above). The temperature gradient across the droplet surface can be as high as $10^\circ$C during the early evaporation stage. After the ethanol component in the binary droplet evaporates, the oscillations stop and the droplet surface temperature slowly rises due to the temperature gradient (the difference between the substrate and ambient temperatures).

%10
\begin{figure}
	\centering
\hspace{0.6cm}  (a) \hspace{5.8cm} (b) \\
\vspace{-0.8cm}
	\includegraphics[width=0.48\textwidth]{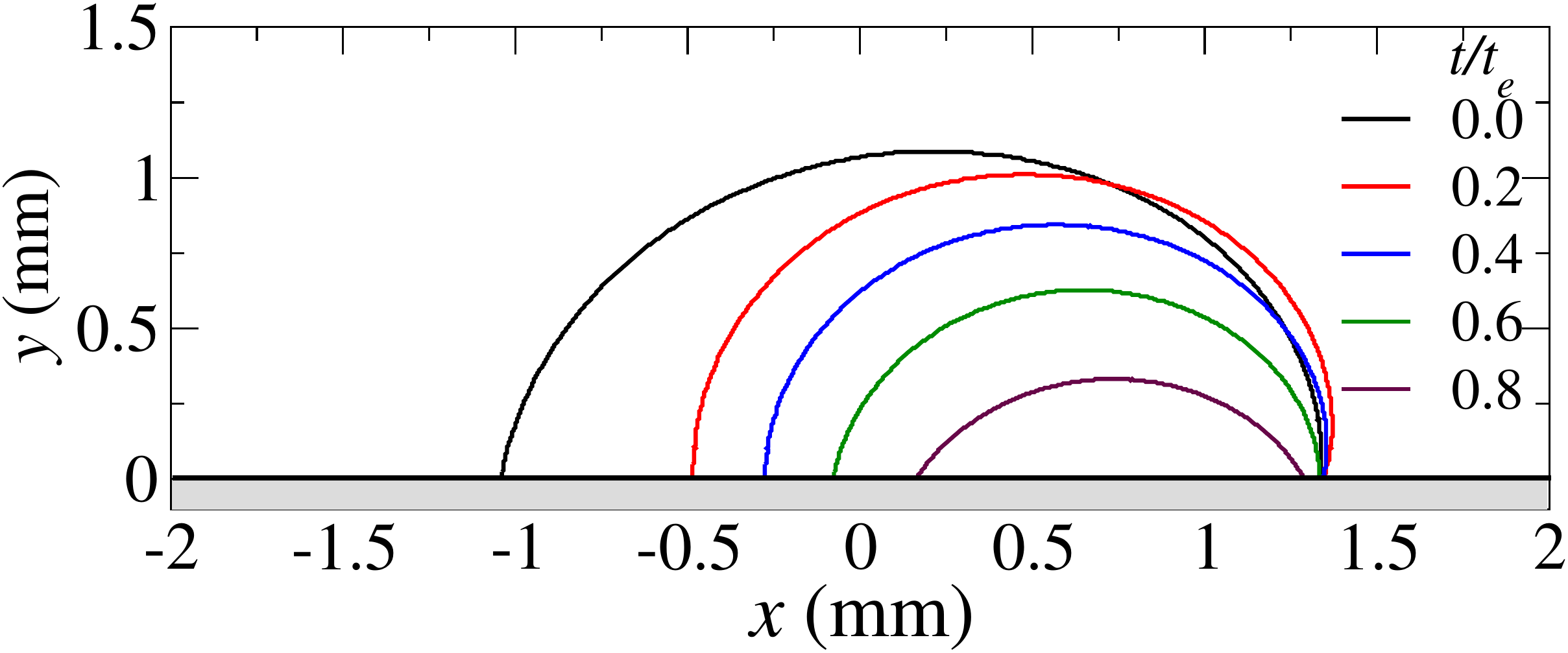} \hspace{2mm} \includegraphics[width=0.48\textwidth]{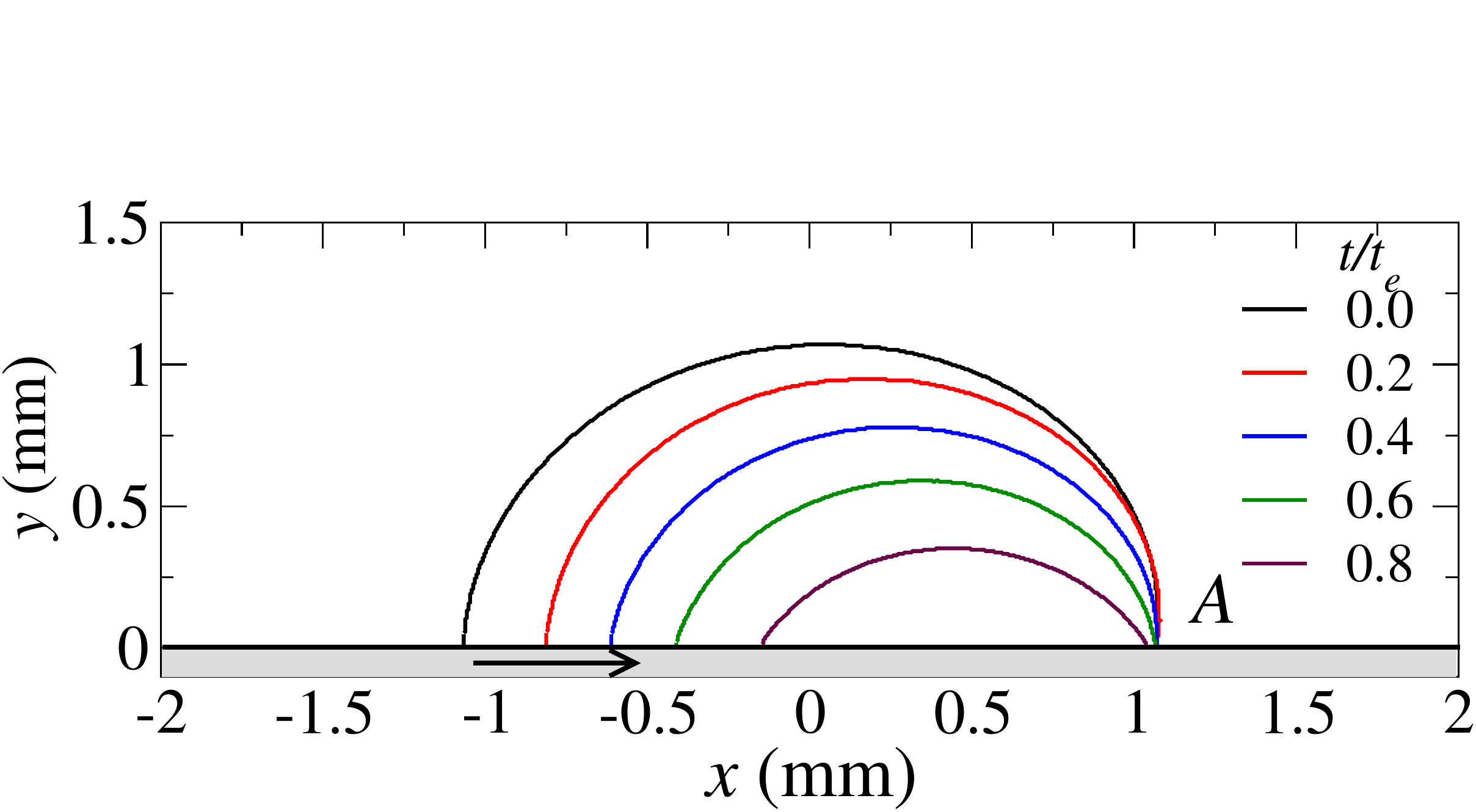}\\
\hspace{0.6cm}  (c) \hspace{5.8cm} (d) \\
	\includegraphics[width=0.48\textwidth]{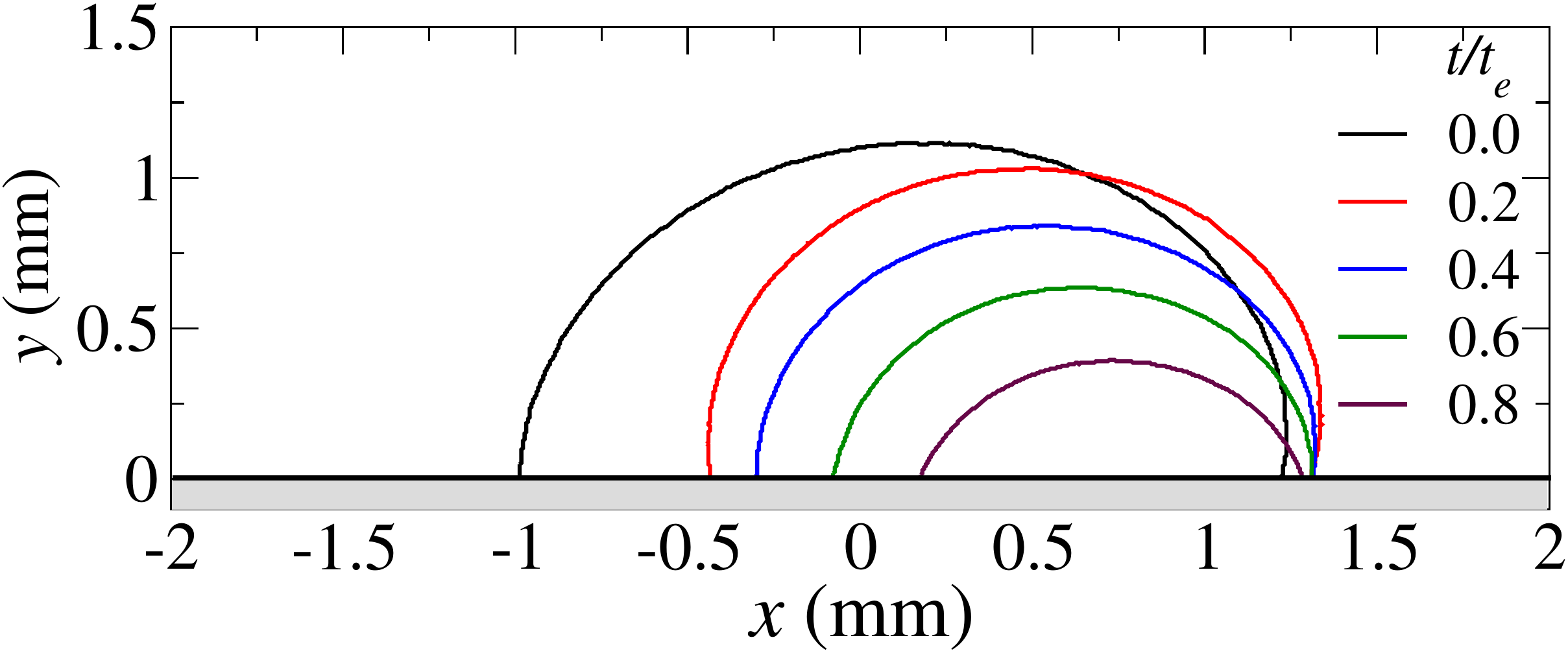} \hspace{2mm} \includegraphics[width=0.48\textwidth]{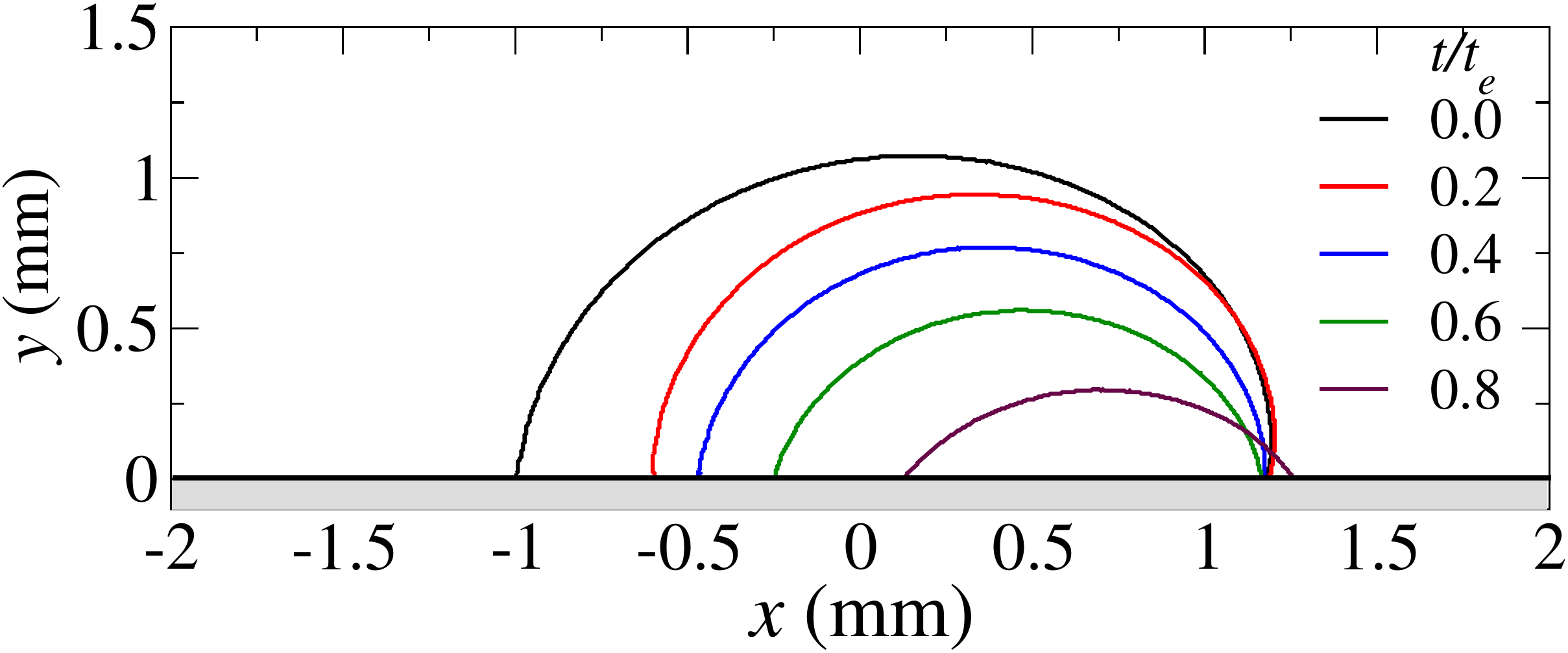} \\
	\caption{Temporal evolutions of the droplet contours at different substrate temperatures at critical angle of inclination. (a) $T_s=40^\circ$C, (b) $T_s=50^\circ$C, (c) $T_s=60^\circ$C and (d) $T_s=70^\circ$C. The arrow in panel (b) indicates the direction in which the droplet moves, and ``A'' means the advancing side of the droplet.}
	\label{fig:Contours_Ts}
\end{figure}

%11
\begin{figure}
\centering
 \hspace{0.6cm}  (a) \hspace{4.5cm} (b) \\
 \includegraphics[width=0.4\textwidth]{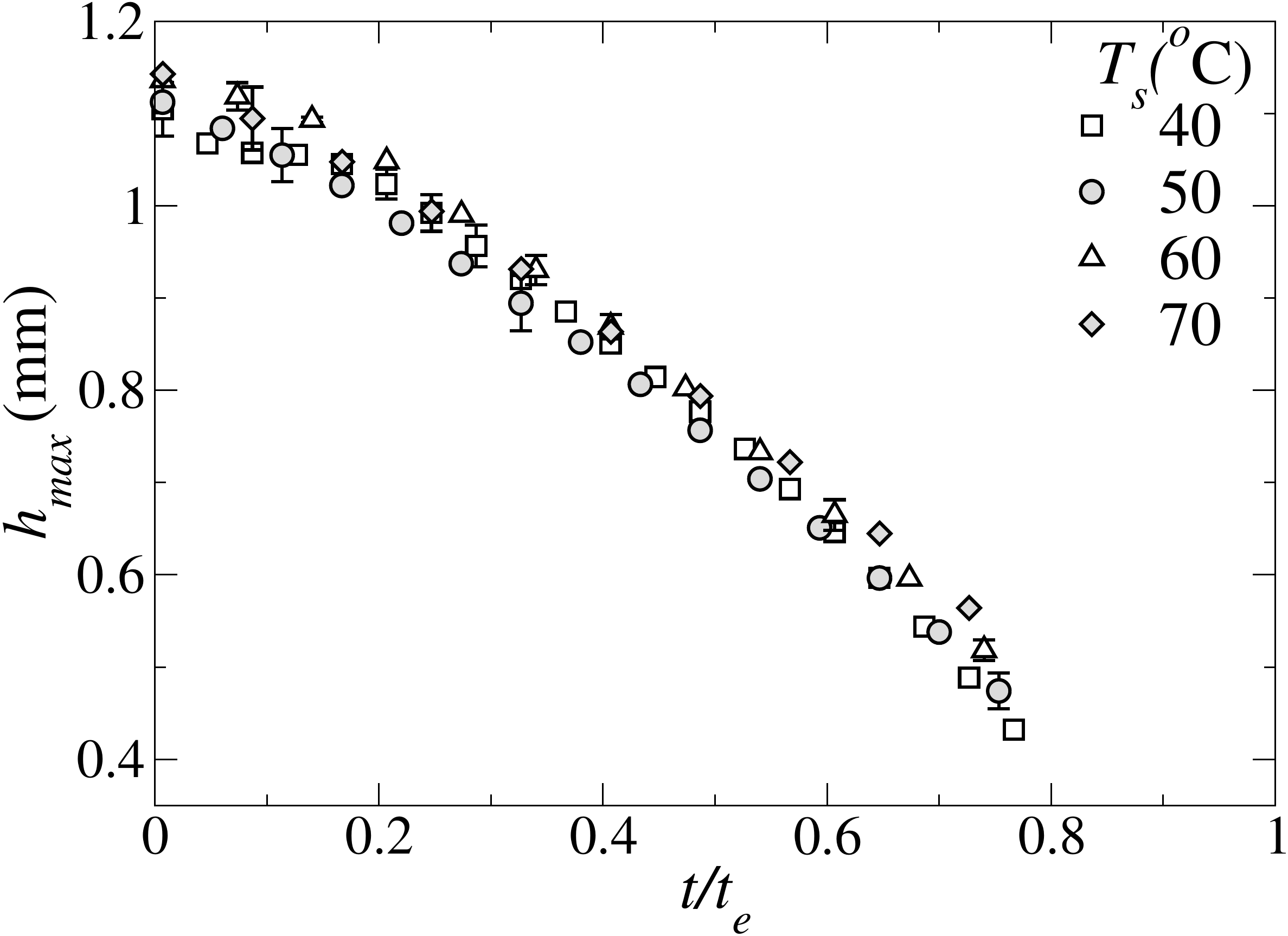} \hspace{2mm} \includegraphics[width=0.4\textwidth]{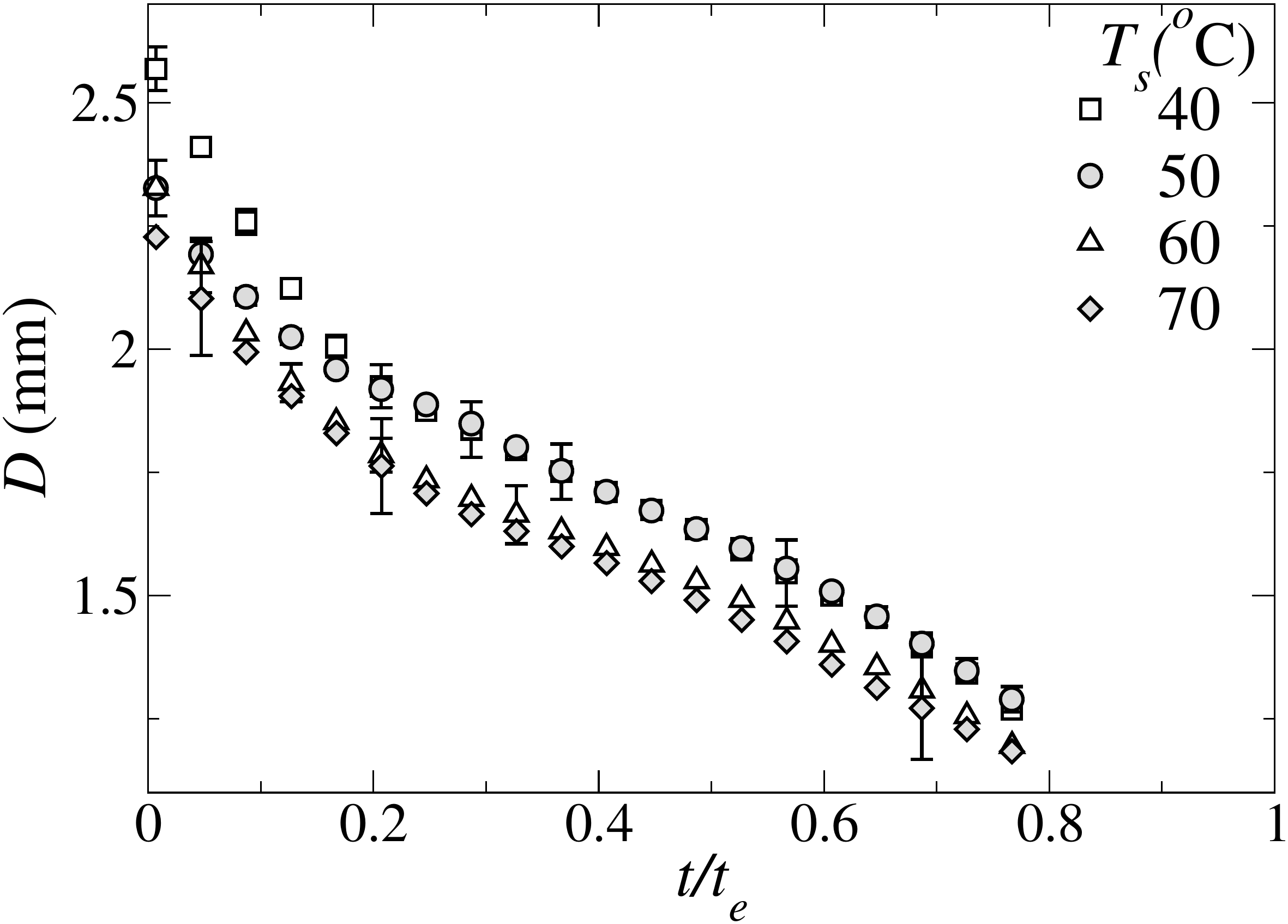}\\
 \hspace{0.6cm}  (c) \hspace{4.5cm} (d) \\
 \includegraphics[width=0.4\textwidth]{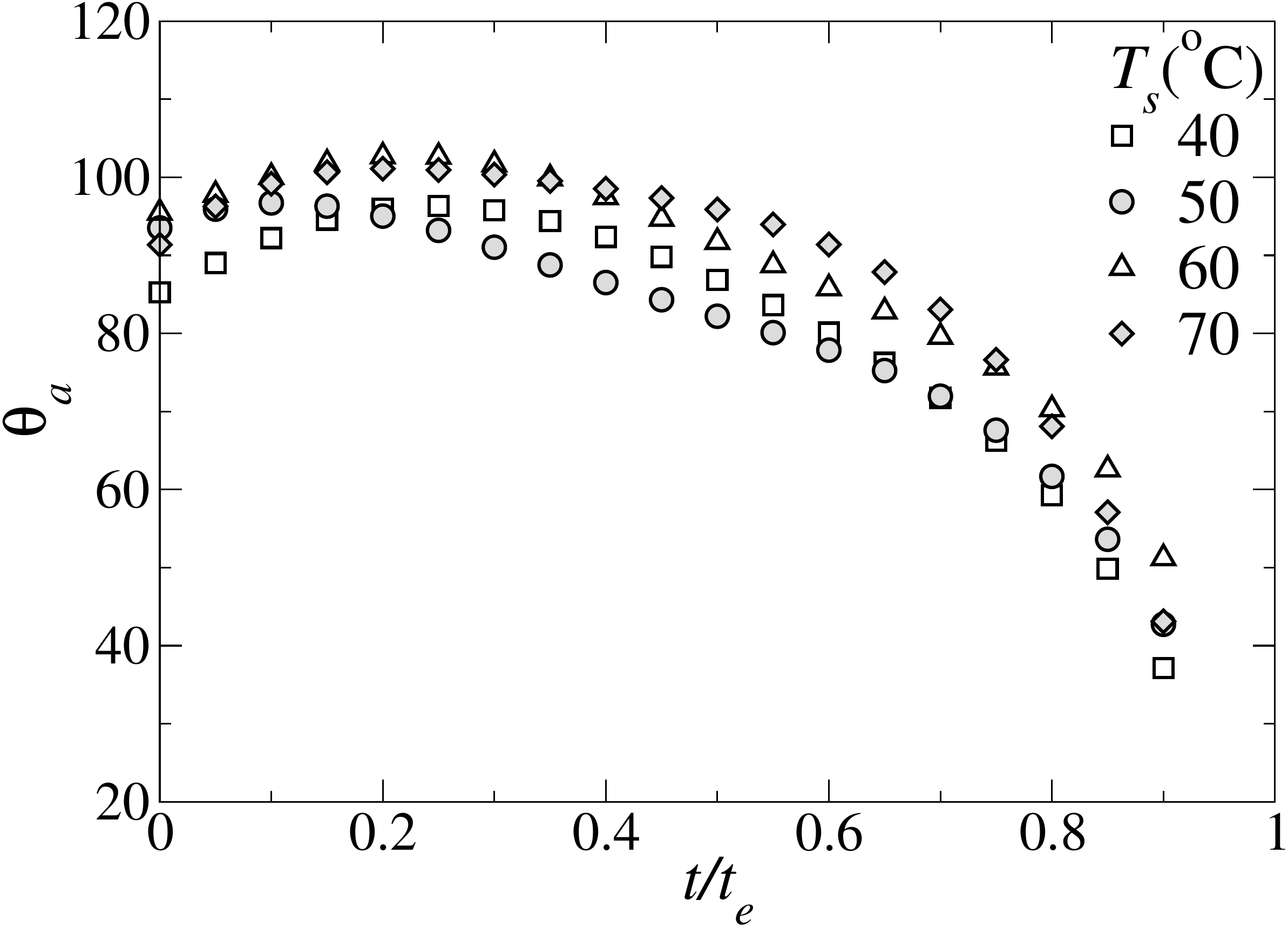} \hspace{2mm} \includegraphics[width=0.4\textwidth]{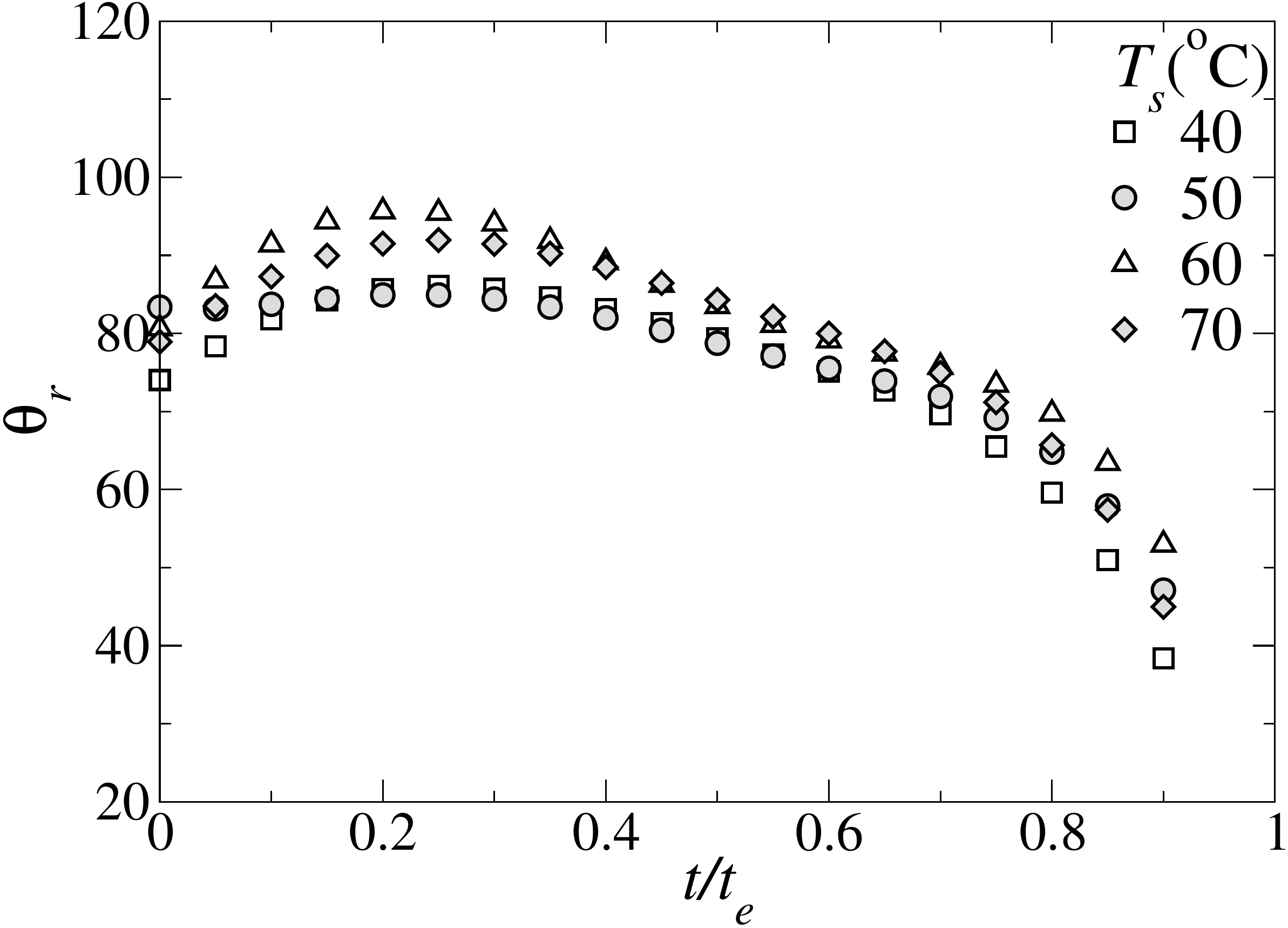} \\
 \hspace{0.5cm} (e) \hspace{4.5cm} (f)\\ 
 \includegraphics[width=0.4\textwidth]{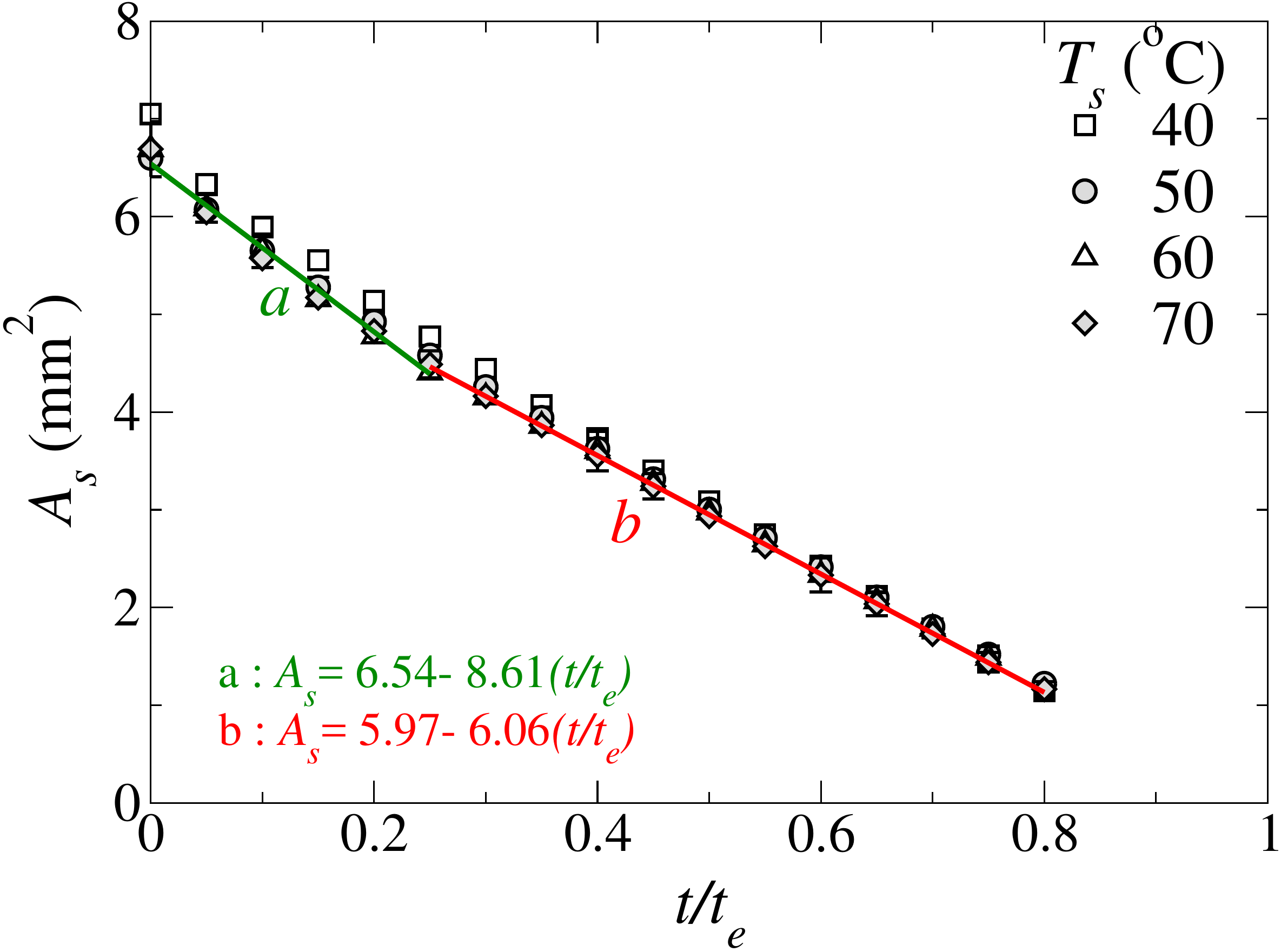} \hspace{2mm} \includegraphics[width=0.4\textwidth]{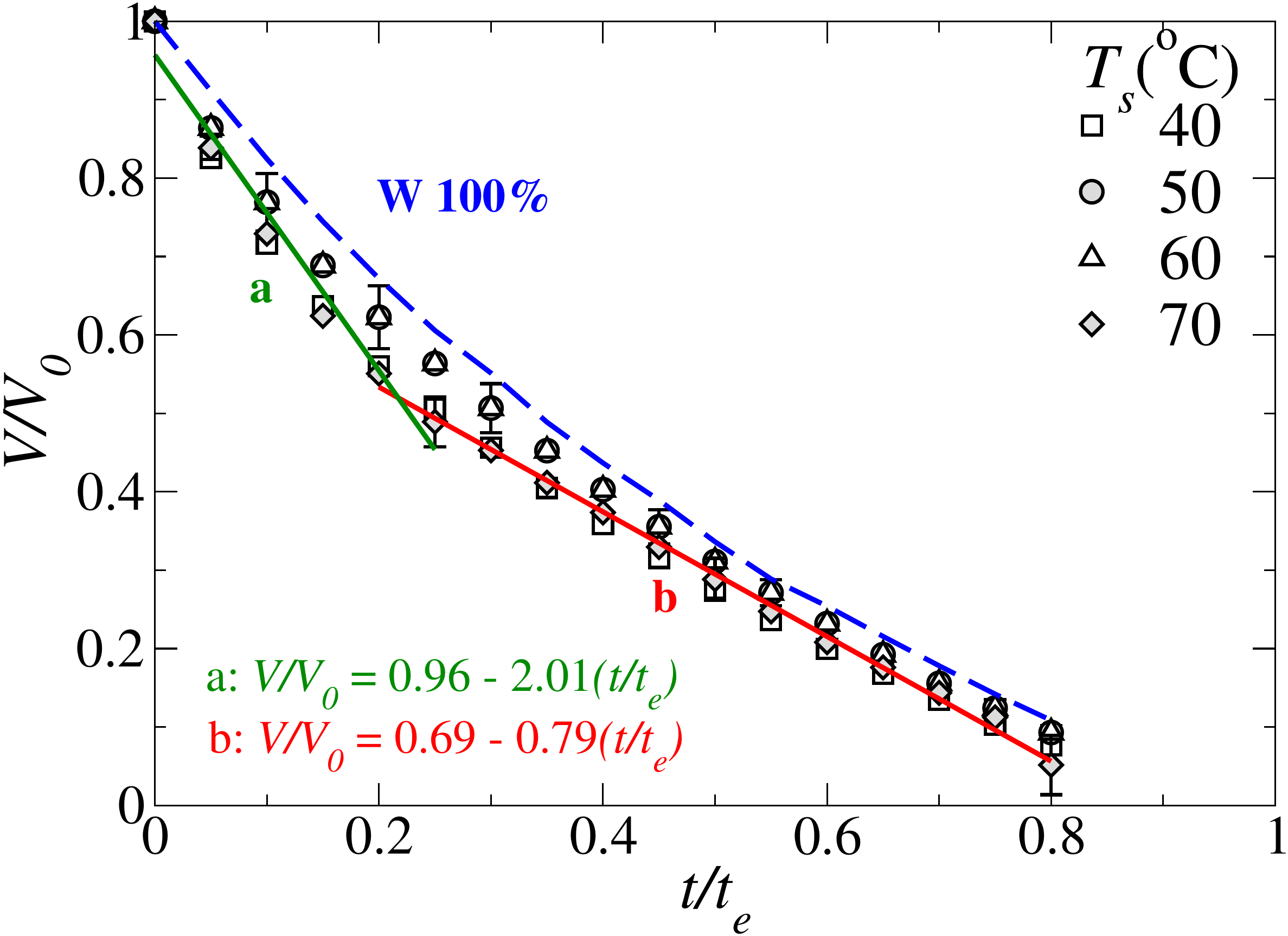}
\caption{Variations of the (a) height ($h$ in mm), (b) wetted diameter of the droplet ($D$ in mm), (c) advancing contact angle ($\theta_a$), (d) receding contact angle ($\theta_r$), (e) surface area ($A_s$ in mm$^2$) with normalised time $\left(t/t_e\right)$ and (f) normalised volume $\left ({V / V_0} \right)$ at different substrate temperatures and the corresponding critical angles as shown in Fig. \ref{fig:geom}(b).}
\label{fig:fig12}
\end{figure}

The temporal evolutions of the side-views of the binary (E 20\% + W 80\%) droplet at different substrate temperatures are presented in Fig. \ref{fig:Side_Ts}. It can be seen that the droplet spreading (wetted radius of the droplet) decreases with the increase in the value of $T_s$, which can be attributed to the increase in the value of the critical angle of inclination with the increase in $T_s$. A close inspection of Fig. \ref{fig:Side_Ts} also reveals that at all values of $T_s$ considered, the binary droplet remains pinned on the advancing side. The temporal evolutions of the contour profile of the droplet at various substrate temperatures reveal the same as shown in Figs. \ref{fig:Contours_Ts}(a)-(d). It can be seen that the droplet indeed remains pinned at the advancing side at low substrate temperatures (see Fig. \ref{fig:Contours_Ts} (a) and (b)), but the advancing contact line of the droplet tends to migrate at high substrate temperatures (see Fig. \ref{fig:Contours_Ts} (c) and (d)) as the droplet evaporates, albeit very slowly. 

Figs. \ref{fig:fig12}(a)-(f) show the variations of the maximum droplet height ($h_{max}$ in mm), the droplet wetted diameter ($D$ in mm), the advancing contact angle $(\theta_a)$, the receding contact angle $(\theta_r)$, the calculated droplet surface area ($A_s$ in mm$^2$) and the calculated droplet volume normalized with the initial volume $(V/V_0)$ plotted against the normalized evaporation time $(t/t_e)$ at different values of the substrate temperature $(T_s)$ and at the corresponding critical inclination angles. {The wetted area and the volume of a typical non-symmetric shaped droplet are calculated in the supporting information using the droplet profiles at different instants during the evaporation.} It can be seen in Fig. \ref{fig:fig12}(a) that the variations of the droplet height are nearly coincident for all values of $T_s$. The trends in the temporal variation of the droplet wetted diameter for all the cases are also similar to one another, with an initially steeper decrease as compared to the later stage for  $t/t_e > 0.2$ (approximately), as can be seen in Fig. \ref{fig:fig12}(b). However, for the entire duration of the evaporation, the wetted diameter is higher for $T_s = 40^\circ$C and $T_s = 50^\circ$C as compared to the other two values of $T_s$. This is probably due to the fact that the critical inclination angle is less at the lower temperatures than that at the higher substrate temperatures. The change in the slope in the temporal variation of the wetted diameter of the droplet can be attributed to the loss of ethanol component from the binary mixture at $t/t_e \approx 0.2$. It is to be noted that the observed decrease in the wetted diameter primarily happens from the receding contact line end, as the advancing end is pinned and does not move.

%12
 \begin{figure}[h]
\centering
 \hspace{0.6cm}  (a) \hspace{4.5cm} (b) \\
 \includegraphics[width=0.4\textwidth]{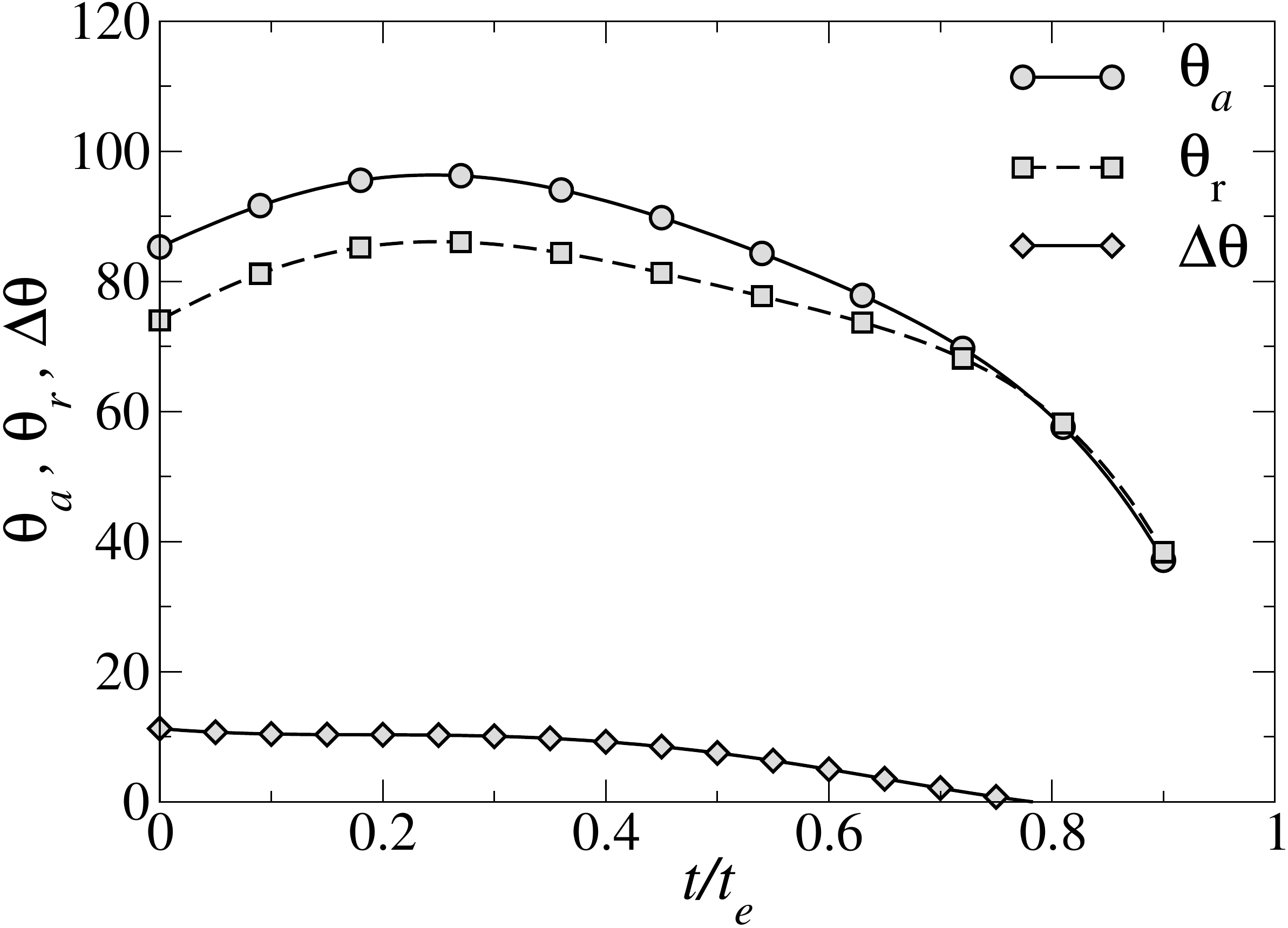} \hspace{2mm} \includegraphics[width=0.4\textwidth]{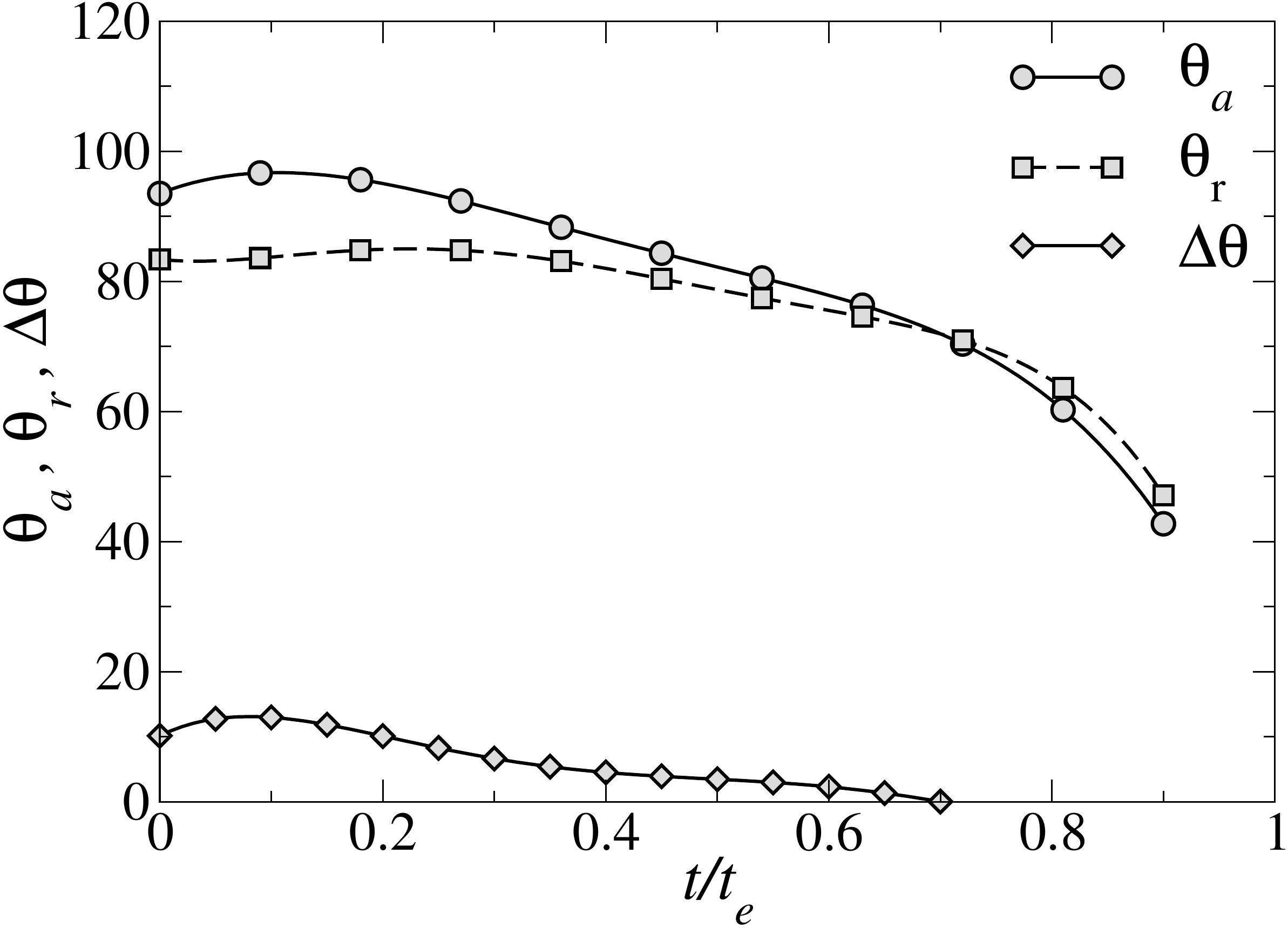}\\
 \hspace{0.6cm}  (c) \hspace{4.5cm} (d) \\
 \includegraphics[width=0.4\textwidth]{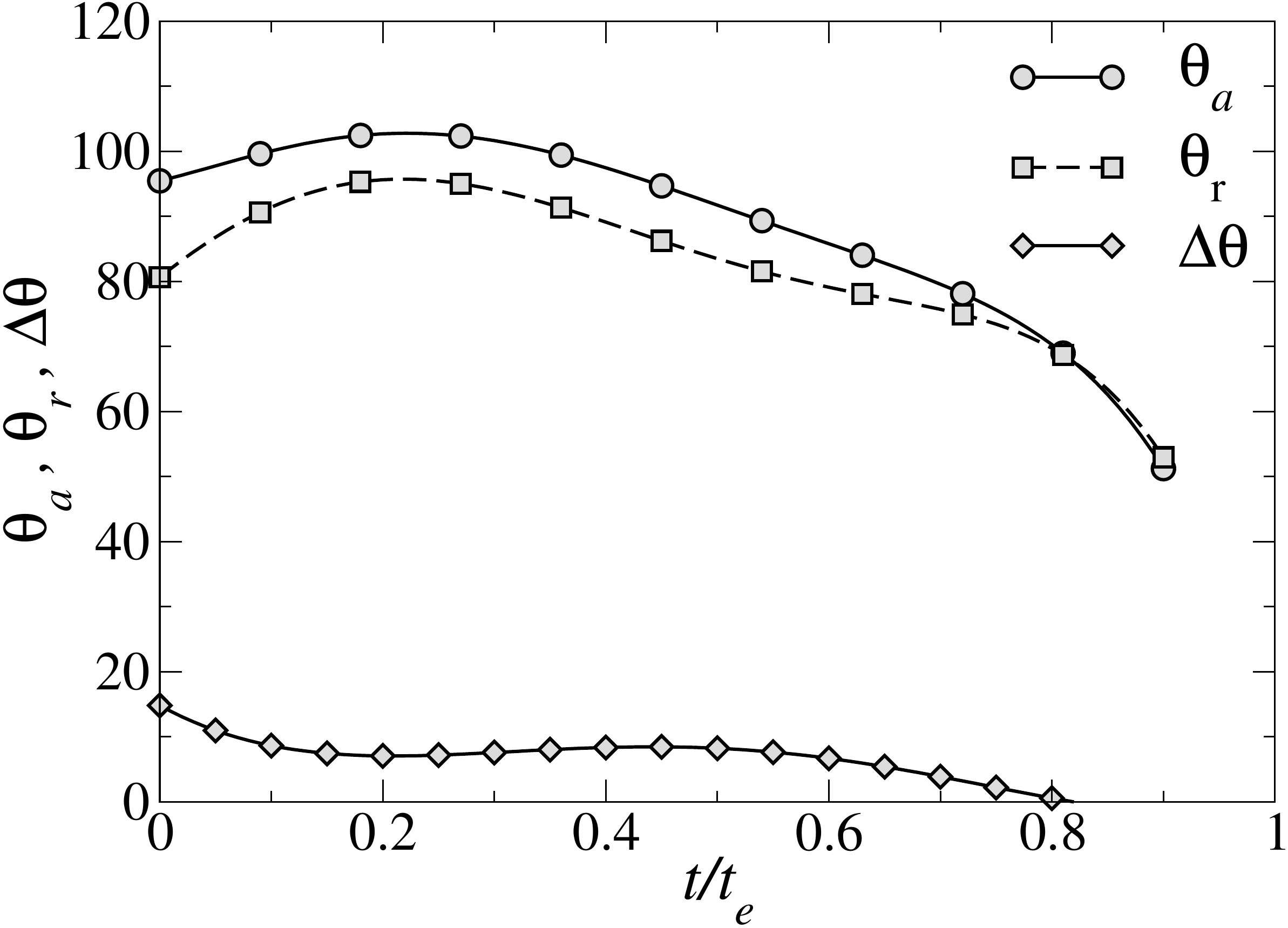} \hspace{2mm} \includegraphics[width=0.4\textwidth]{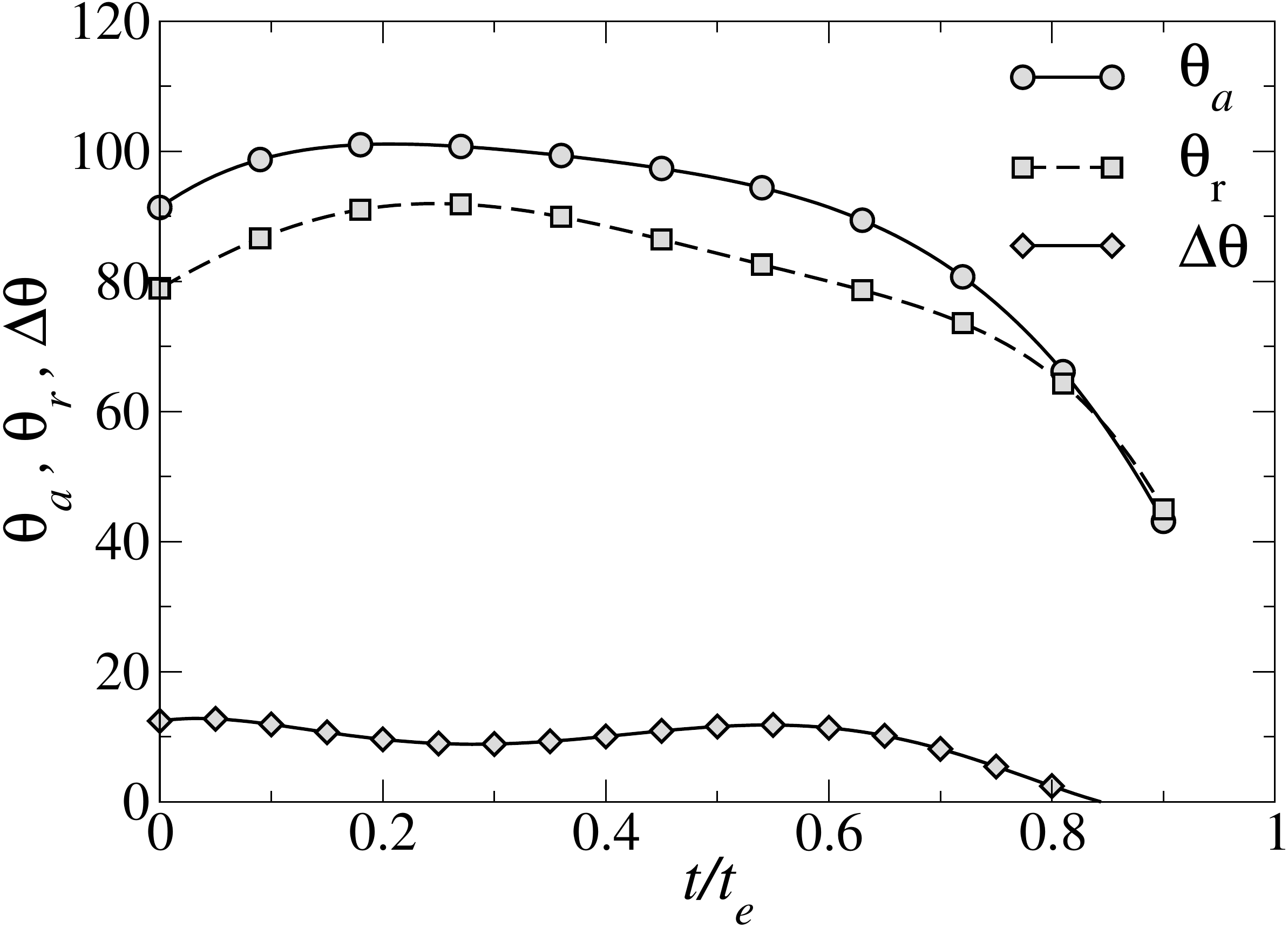} \\
\caption{Temporal variations of the advancing contact angle $(\theta_a)$, receding contact angle $(\theta_r)$ and contact angle hysteresis $(\Delta \theta_a)$ at (a) $T_s=40^\circ$C, (b) $T_s=50^\circ$C, (c) $T_s=60^\circ$C, (d) $T_s=70^\circ$C at their corresponding critical angles as shown in Fig. \ref{fig:geom}(b).}
\label{fig:fig8}
\end{figure}

It can be seen in Figs. \ref{fig:fig12}(c)-(d) that in consistent with the larger wetted diameter for $40^\circ$C and $50^\circ$C cases, the advancing and the receding contact angles are also lesser than the corresponding values for the $60^\circ$C and the $70^\circ$C cases. Three distinct stages in the contact line dynamics can be clearly visible in Figs. \ref{fig:fig12}(c)-(d). It can be observed that in $0<t/t_e<0.2$ (early stage), the advancing and the receding contact angles show a slow increase in their absolute values as the percentage of ethanol decreases and the droplet tries to equilibrate to the contact angle values of the pure water droplet. Between $0.2<t/t_e<0.8$ (intermediate stage), the droplet contact angles exhibit a slow decrease (due to the pure water component) as the droplet evaporates steadily. At the last stage of the evaporation, the contact angle value drops off rapidly as the small residual droplet undergoes a rapid unsteady contraction. The three distinct stages of the dynamics of the contact angles may be modeled with individual linear fits, even though the fitting parameters differ for different substrate temperatures {(Fig. 17 and Table 1 in supporting information)}. 

For all values of $T_s$, it can also be seen that the advancing contact angle is higher than the receding contact angle during $0<t/t_e<0.7$. This behavior is clearly demonstrated in Fig. \ref{fig:fig8} which presents the advancing and receding contact angle profiles of each case in one figure, as well as the difference between the two angles with the normalized time. At the late stage of evaporation, the droplet is too small for gravity-induced asymmetry to be substantial, and thus the difference between the advancing and the receding contact angles (contact angle hysteresis) decreases to zero. Figs. \ref{fig:fig12}(e) and (f) show the variations of the droplet surface area $(A_s)$ and normalised droplet volume $(V/V_0)$ with normalized time $(t/t_e)$ for different substrate temperature, $T_s$. It can be seen that the variations collapse to a single curve at all values of $T_s$. Furthermore, the variations of $A_s$ and $V/V_0$ exhibit two distinct nearly linear regions, namely early ethanol evaporation dominated region ($0<t/t_e<0.2$) and late water dominated evaporation region ($0.2<t/t_e<0.8$). Thus the profile can be modeled using two linear fits for these two evaporation regions with an inflection point in between. For the configuration considered in the present study, the linear fits in $A_s$ are given by $A_s=6.54 - 8.61 (t/t_e)$ and $A_s=5.97 - 6.06 (t/t_e)$. Similarly, the linear fits in $V/V_0$ are given by $V/V_0=0.96 - 2.01 (t/t_e)$ and $V/V_0=0.69  - 0.79 (t/t_e)$, respectively. These linear fits are also shown in Figs. \ref{fig:fig12}(e) and (f). In panel (f), the result for a pure water droplet has also been added to compare with that of the binary mixture. {Fig. 19 (supporting information) shows that the linear fits match with the experiment based volume data very well.} Thus, despite the complex thermo-solutal Marangoni-force driven convection observed for the binary droplet using the thermal camera, the suitably normalized volume against evaporation time profiles show very simple universal behavior at all substrate temperatures at their corresponding critical angles.

\begin{figure}
\centering
 \hspace{0.6cm}  (a) \hspace{4.5cm} (b) \\
 \includegraphics[width=0.4\textwidth]{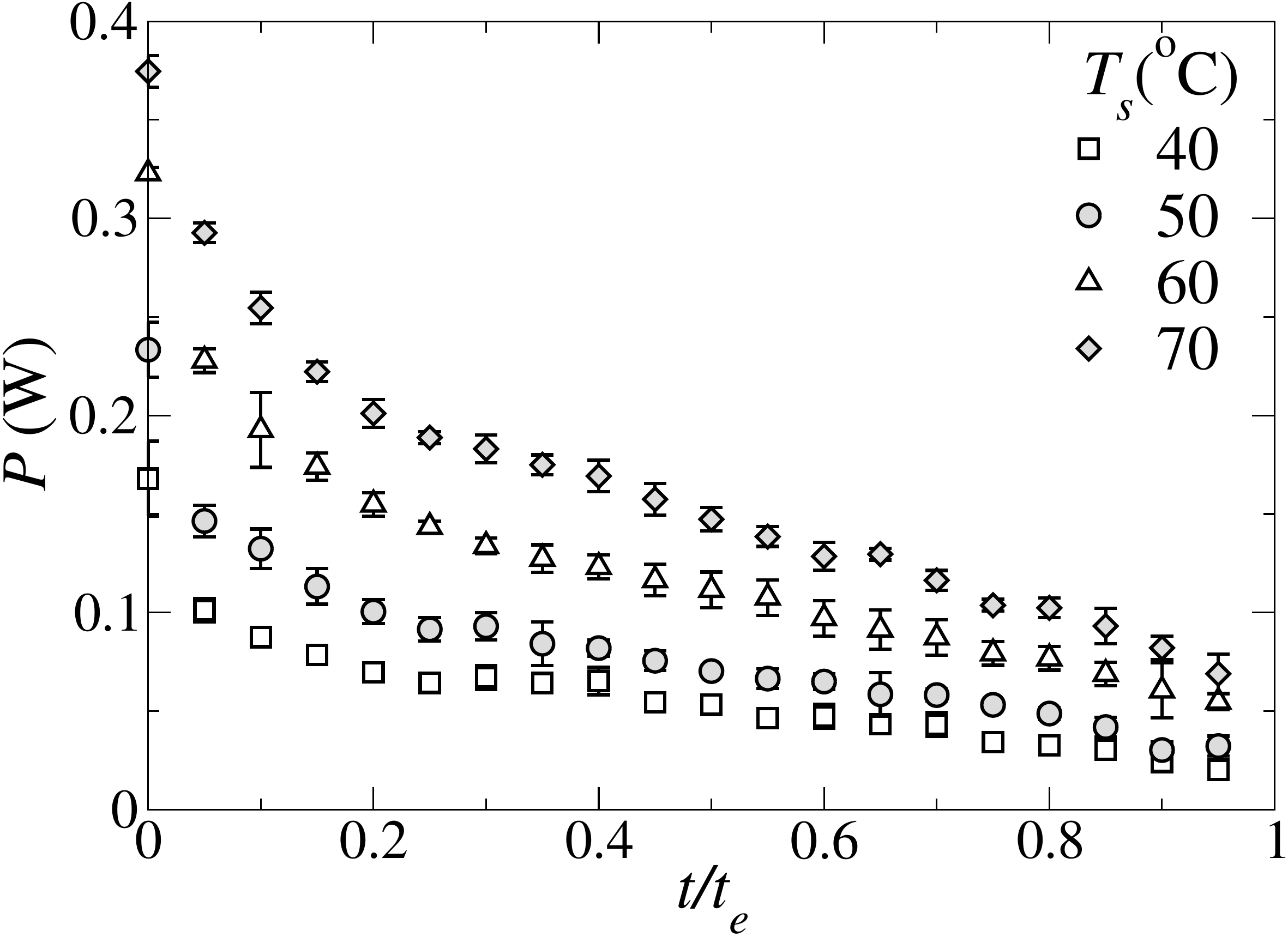} \hspace{2mm} \includegraphics[width=0.4\textwidth]{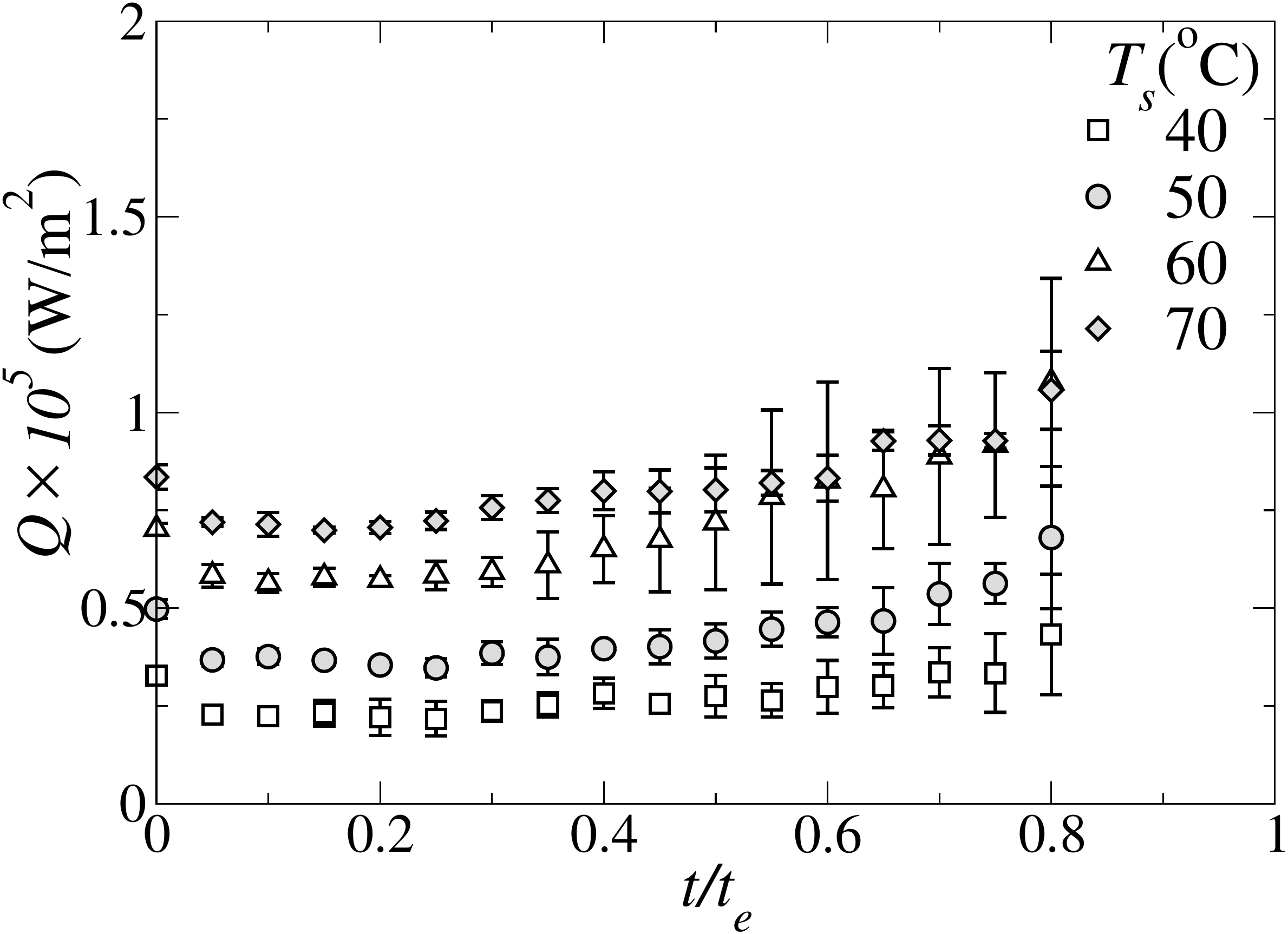}\\
 \caption{Variations of the (a) amount of heat supplied to the drop and (b) the corresponding heat flux with normalised time $(t/t_e)$ at different substrate temperatures.}
\label{fig:fig10}
\end{figure}

Fig. \ref{fig:fig10} shows the variations of the total heat transfer rate ($P$ in Watts) and the heat flux ($Q$ in $W/m^2$) absorbed from the substrate area wetted by the droplet as a function of the normalized time, $t/t_e$ at different values of the substrate temperature, $T_s$. The heat flow absorbed by the drop is calculated similarly as presented by Sobac {\it et al} \cite{sobac2012thermocapillary}. As expected, the heat transfer rate as well as the heat flux into the droplet increase with an increase in the substrate temperature as the temperature difference between the initially deposited droplet and the substrate temperature increases. The absolute value of the Wattage decreases with time at each value of $T_s$ since the wetted area shrinks (\ref{fig:fig12}b) as the droplet evaporates. However, the heat flux value (i.e. the heat absorbed per unit wetted surface area per unit normalized time) remains relatively constant at the early times and increases slightly near the end stage of evaporation. This may be attributed to the decrease in the droplet contact angles over time as shown in \ref{fig:fig8} which in turn leads to an increase in the droplet surface area to volume ratio ($A_s/V$), thereby enhancing the evaporation rate. It must be noted, however, that the error bars associated with the heat flux values increase at the late stage, as the smaller wetted area creates larger uncertainty in the flux calculation.

%13

%\clearpage

\section{Conclusions}
\label{sec:conc}

The evaporation dynamics of a binary (E 20\% + W 80\%) sessile droplet of 5 $\mu$l placed on an inclined substrate at just below the critical angles for different substrate temperatures is investigated experimentally. A customized goniometer equipped with a CMOS camera and an infrared camera is used to study the evaporation dynamics and the associated thermo-solutal Marangoni convection at the liquid-air interface. The inclined substrate consists of polytetrafluoroethylene (PTFE) tape of thickness 100 $\mu$m pasted on an aluminum plate coated with black paint to minimize the reflection from the IR camera. A PID controller is used to maintain the substrate at different temperatures, namely, $40^\circ$C, $50^\circ$C, $60^\circ$C and $70^\circ$C. The experiments are performed at a constant ambient temperature of $21^\circ$C and relative humidity of $70 \pm 10$ \% in an air-conditioned room. We found that the critical angle of inclination at which the binary droplet starts to migrate under the action of gravity increases with increasing the substrate temperature; however, a pure-water droplet remains stationary at all inclinations . Thus, all the experiments are performed just below the critical angle of inclination for the binary droplet at the corresponding substrate temperature. Due to the asymmetry in the advancing and receding contact angles, the evaporation dynamics is found to be different from that observed in the case of a flat horizontal substrate. To the best of our knowledge, the evaporation dynamics of a binary droplet at the critical angles for different substrate temperatures has been investigated for the first time in the present study. 

Comparison of the evaporation dynamics of pure-water and binary (E 20\% + W 80\%) sessile droplets at an elevated substrate temperature ($T_s=70^\circ$C) and $\alpha=45^\circ$ reveals that most of the early evaporation occurs at the receding contact angle side for both the liquids. The droplet is observed to remain pinned in the advancing side during the evaporation process, while the receding side contracts. In the case of binary droplet, as water is heavier and less volatile than ethanol, it moves to the advancing side due to gravity and the thermo-solutal Marangoni force causing the ethanol component to accumulate and evaporate faster at the receding side of the droplet. In contrast to the pure-water droplet, the binary droplet exhibits two distinct oscillatory water-rich cold regions of different sizes around the advancing contact line during the early stage of evaporation. The residual ethanol, accumulated as a hot and rapidly evaporating cell in the receding contact line region, collapses at the end of the early stage to produce thermal pulsations that promotes mixing between the cold and hot regions. It is observed that the oscillatory thermo-solutal convection cells and thermal pulsation phenomenon are more intense and long lasting (with respect to the total evaporation time) at higher substrate temperatures. At the later stage, the cold regions combine after the evaporation of the more volatile ethanol component, and the thermal patterns become similar to the pure-water droplet. It is also found that the spreading of the droplet decreases with increasing substrate temperature. 

Three distinct stages in the contact line dynamics are observed at all the values of the substrate temperature considered in the present study. They are, namely, an early stage when the advancing and the receding contact angles increase, an intermediate stage when the droplet contact angles exhibit a slow decrease and a late stage when both contact angles decrease rapidly as the small residual droplet undergoes a rapid unsteady contraction. Despite the complexity in convection dynamics, the evaporation rate exhibits a universal linear behavior in the normalized time at different substrate temperatures which can be represented by piecewise linear fits at the early (ethanol dominated) and late (water dominated) stages of evaporation. The total heat transfer rate and the heat flux absorbed from the wetted area have also been calculated at different substrate temperatures and it is found that increasing the substrate temperature increases the heat transfer rate as well as the heat flux to the droplet.\\

\noindent{\bf Acknowledgement:} {K. C. S. thanks Science \& Engineering Research Board, India for their financial support (Grant number: MTR/2017/000029). }

\section*{References}
%\bibliography{bibl}

\begin{thebibliography}{10}
\expandafter\ifx\csname url\endcsname\relax
  \def\url#1{\texttt{#1}}\fi
\expandafter\ifx\csname urlprefix\endcsname\relax\def\urlprefix{URL }\fi
\expandafter\ifx\csname href\endcsname\relax
  \def\href#1#2{#2} \def\path#1{#1}\fi

\bibitem{lim2012deposit}
T.~Lim, J.~Yang, S.~Lee, J.~Chung, D.~Hong, Deposit pattern of inkjet printed
  pico-liter droplet, Int. J. Pr. Eng. Man. 13~(6) (2012) 827--833.

\bibitem{kim2006direct}
D.~Kim, S.~Jeong, B.~K. Park, J.~Moon, Direct writing of silver conductive
  patterns: Improvement of film morphology and conductance by controlling
  solvent compositions, Appl. Phys. Lett. 89~(26) (2006) 264101.

\bibitem{park2006control}
J.~Park, J.~Moon, Control of colloidal particle deposit patterns within
  picoliter droplets ejected by ink-jet printing, Langmuir 22~(8) (2006)
  3506--3513.

\bibitem{lim2009experimental}
T.~Lim, S.~Han, J.~Chung, J.~T. Chung, S.~Ko, C.~P. Grigoropoulos, Experimental
  study on spreading and evaporation of inkjet printed pico-liter droplet on a
  heated substrate, Int. J. Heat Mass Transf. 52~(1-2) (2009) 431--441.

\bibitem{koo2006fabrication}
H.~Koo, M.~Chen, P.~Pan, L.~Chou, F.~Wu, S.~Chang, T.~Kawai, Fabrication and
  chromatic characteristics of the greenish lcd colour-filter layer with
  nano-particle ink using inkjet printing technique, Displays 27~(3) (2006)
  124--129.

\bibitem{tekin2004ink}
E.~Tekin, B.~J. de~Gans, U.~S. Schubert, Ink-jet printing of polymers--from
  single dots to thin film libraries, Journal of Materials Chemistry 14~(17)
  (2004) 2627--2632.

\bibitem{de2004inkjet}
B.~J. de~Gans, U.~S. Schubert, Inkjet printing of well-defined polymer dots and
  arrays, Langmuir 20~(18) (2004) 7789--7793.

\bibitem{deegan1997capillary}
R.~D. Deegan, O.~Bakajin, T.~F. Dupont, G.~Huber, S.~R. Nagel, T.~A. Witten,
  Capillary flow as the cause of ring stains from dried liquid drops, Nature
  389~(6653) (1997) 827.

\bibitem{yanagisawa2014investigation}
K.~Yanagisawa, M.~Sakai, T.~Isobe, S.~Matsushita, A.~Nakajima, Investigation of
  droplet jumping on superhydrophobic coatings during dew condensation by the
  observation from two directions, Appl. Surf. Sci. 315 (2014) 212--221.

\bibitem{tripathi2015evaporating}
M.~K. Tripathi, K.~C. Sahu, Evaporating falling drop, Procedia IUTAM 15 (2015)
  201--206.

\bibitem{langmuir1918evaporation}
I.~Langmuir, The evaporation of small spheres, Phys. Rev. 12~(5) (1918) 368.

\bibitem{picknett1977evaporation}
R.~G. Picknett, R.~Bexon, The evaporation of sessile or pendant drops in still
  air, J. Colloid Interface Sci. 61~(2) (1977) 336--350.

\bibitem{sobac2012thermal}
B.~Sobac, D.~Brutin, Thermal effects of the substrate on water droplet
  evaporation, Phys. Rev. E. 86~(2) (2012) 021602.

\bibitem{birdi1989study}
K.~S. Birdi, D.~T. Vu, A.~Winter, A study of the evaporation rates of small
  water drops placed on a solid surface, J. Phys. Chem. 93~(9) (1989)
  3702--3703.

\bibitem{sobac2012thermocapillary}
B.~Sobac, D.~Brutin, Thermocapillary instabilities in an evaporating drop
  deposited onto a heated substrate, Phys. Fluids 24~(3) (2012) 032103.

\bibitem{brutin2018recent}
D.~Brutin, V.~Starov, Recent advances in droplet wetting and evaporation, Chem.
  Soc. Rev. 47~(2) (2018) 558--585.

\bibitem{extrand1995liquid}
C.~W. Extrand, Y.~Kumagai, Liquid drops on an inclined plane: the relation
  between contact angles, drop shape, and retentive force, J. Colloid Interf.
  Sci. 170~(2) (1995) 515--521.

\bibitem{kim2002sliding}
H.~Kim, H.~J. Lee, B.~H. Kang, Sliding of liquid drops down an inclined solid
  surface, J. Colloid Interf. Sci. 247~(2) (2002) 372--380.

\bibitem{chou2012drops}
T.-H. Chou, S.-J. Hong, Y.-J. Sheng, H.-K. Tsao, Drops sitting on a tilted
  plate: receding and advancing pinning, Langmuir 28~(11) (2012) 5158--5166.

\bibitem{kim2017evaporation}
J.~Y. Kim, I.~G. Hwang, B.~M. Weon, Evaporation of inclined water droplets,
  Sci. Rep. 7 (2017) 42848.

\bibitem{annapragada2012droplet}
S.~R. Annapragada, J.~Y. Murthy, S.~V. Garimella, Droplet retention on an
  incline, Int. J. Heat Mass Transf. 55~(5-6) (2012) 1457--1465.

\bibitem{yilbas2017dynamics}
B.~S. Yilbas, A.~Al-Sharafi, H.~Ali, N.~Al-Aqeeli, Dynamics of a water droplet
  on a hydrophobic inclined surface: influence of droplet size and surface
  inclination angle on droplet rolling, RSC Adv. 7~(77) (2017) 48806--48818.

\bibitem{ellaban2017instability}
E.~Ellaban, J.~Pascal, S.~D’Alessio, Instability of a binary liquid film
  flowing down a slippery heated plate, Phys. Fluids 29~(9) (2017) 092105.

\bibitem{sefiane2003experimental}
K.~Sefiane, L.~Tadrist, M.~Douglas, Experimental study of evaporating
  water--ethanol mixture sessile drop: influence of concentration, Int. J. Heat
  Mass Transf. 46~(23) (2003) 4527--4534.

\bibitem{christy2011flow}
J.~R.~E. Christy, Y.~Hamamoto, K.~Sefiane, Flow transition within an
  evaporating binary mixture sessile drop, Phys. Rev. Lett. 106~(20) (2011)
  205701.

\bibitem{bennacer2014vortices}
R.~Bennacer, K.~Sefiane, Vortices, dissipation and flow transition in volatile
  binary drops, J. Fluid Mech. 749 (2014) 649--665.

\bibitem{ozturk2018evaporation}
T.~Ozturk, H.~Y. Erbil, Evaporation of water-ethanol binary sessile drop on
  fluoropolymer surfaces: Influence of relative humidity, Colloids Surf. A 553
  (2018) 327--336.

\bibitem{schofield2018lifetimes}
F.~G.~H. Schofield, S.~K. Wilson, D.~Pritchard, K.~Sefiane, The lifetimes of
  evaporating sessile droplets are significantly extended by strong thermal
  effects, J. Fluid Mech. 851 (2018) 231--244.

\bibitem{feng2017octagon}
H.~Feng, K.~S.-L. Chong, K.~S. Ong, F.~Duan, Octagon to square wetting area
  transition of water--ethanol droplets on a micropyramid substrate by
  increasing ethanol concentration, Langmuir 33~(5) (2017) 1147--1154.

\bibitem{shi2009wetting}
L.~Shi, P.~Shen, D.~Zhang, Q.~Lin, Q.~Jiang, Wetting and evaporation behaviors
  of water--ethanol sessile drops on ptfe surfaces, Surf. Interface Anal.
  41~(12-13) (2009) 951--955.

\bibitem{sefiane2008wetting}
K.~Sefiane, S.~David, M.~Shanahan, Wetting and evaporation of binary mixture
  drops, J. Phys. Chem. B. 112~(36) (2008) 11317--11323.

\bibitem{wang2008evaporation}
Z.~Wang, X.-F. Peng, A.~S. Mujumdar, A.~Su, D.-J. Lee, Evaporation of
  ethanol-water mixture drop on horizontal substrate, Dry. Technol. 26~(6)
  (2008) 806--810.

\bibitem{cheng2006evaporation}
A.~K.~H. Cheng, D.~M. Soolaman, H.-Z. Yu, Evaporation of microdroplets of
  ethanol- water mixtures on gold surfaces modified with self-assembled
  monolayers, J. Phys. Chem. B. 110~(23) (2006) 11267--11271.

\bibitem{karpitschka2017marangoni}
S.~Karpitschka, F.~Liebig, H.~Riegler, Marangoni contraction of evaporating
  sessile droplets of binary mixtures, Langmuir 33~(19) (2017) 4682--4687.

\bibitem{karapetsas2016evaporation}
G.~Karapetsas, K.~C. Sahu, O.~K. Matar, Evaporation of sessile droplets laden
  with particles and insoluble surfactants, Langmuir 32~(27) (2016) 6871--6881.

\bibitem{karapetsas2014thermocapillary}
G.~Karapetsas, K.~C. Sahu, K.~Sefiane, O.~K. Matar, Thermocapillary-driven
  motion of a sessile drop: effect of non-monotonic dependence of surface
  tension on temperature, Langmuir 30~(15) (2014) 4310--4321.

\bibitem{gurrala2019evaporation}
P.~Gurrala, P.~Katre, S.~Balusamy, S.~Banerjee, K.~C. Sahu, Evaporation of
  ethanol-water sessile droplet of different compositions at an elevated
  substrate temperature, Int. J. Heat Mass Transf. 145 (2019) 118770.

\bibitem{yonemoto2018sliding}
Y.~Yonemoto, S.~Suzuki, S.~Uenomachi, T.~Kunugi, Sliding behaviour of
  water-ethanol mixture droplets on inclined low-surface-energy solid, Int. J.
  Heat Mass Transf. 120 (2018) 1315--1324.

\bibitem{innocenzi2008evaporation}
P.~Innocenzi, L.~Malfatti, S.~Costacurta, T.~Kidchob, M.~Piccinini,
  A.~Marcelli, Evaporation of ethanol and ethanol- water mixtures studied by
  time-resolved infrared spectroscopy, J. Phys. Chem. A 112~(29) (2008)
  6512--6516.

\bibitem{saenz2017dynamics}
P.~J. S{\'a}enz, A.~W. Wray, Z.~Che, O.~K. Matar, P.~Valluri, J.~Kim,
  K.~Sefiane, Dynamics and universal scaling law in geometrically-controlled
  sessile drop evaporation, Nat. Commun. 8 (2017) 14783.

\bibitem{mamalis2016motion}
D.~Mamalis, V.~Koutsos, K.~Sefiane, On the motion of a sessile drop on an
  incline: Effect of non-monotonic thermocapillary stresses, Appl. Phys. Lett.
  109~(23) (2016) 231601.

\end{thebibliography}

\clearpage

\section*{Supporting information}

\begin{figure}[h]
\centering
 \includegraphics[width=0.7\textwidth]{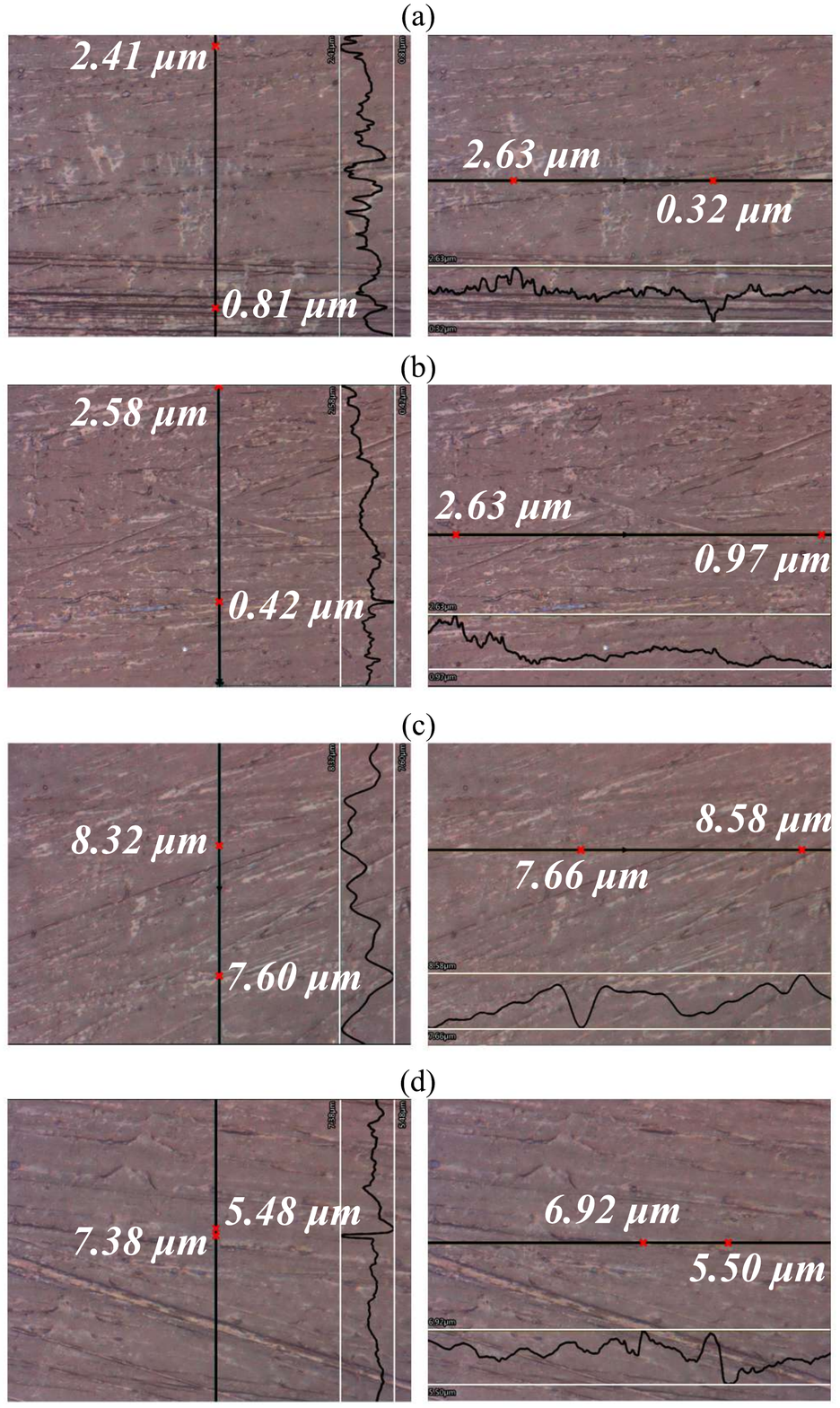}
\caption{Optical microscopy images of the PTFE substrate at (a) $T_s =40 ^{\circ}$C, (b) $T_s =50 ^{\circ}$C, (c) $T_s =60 ^{\circ}$C, and (d) $T_s =70 ^{\circ}$C. The images are taken at different locations on the tape and it can be observed that the tape is thermally stable at different temperatures considered. The minimum and maximum values of the roughness (presented in between two white lines) are written at two  locations.}
\label{fig:figS1}
\end{figure}

\begin{figure}
\centering
\includegraphics[width=0.5\textwidth]{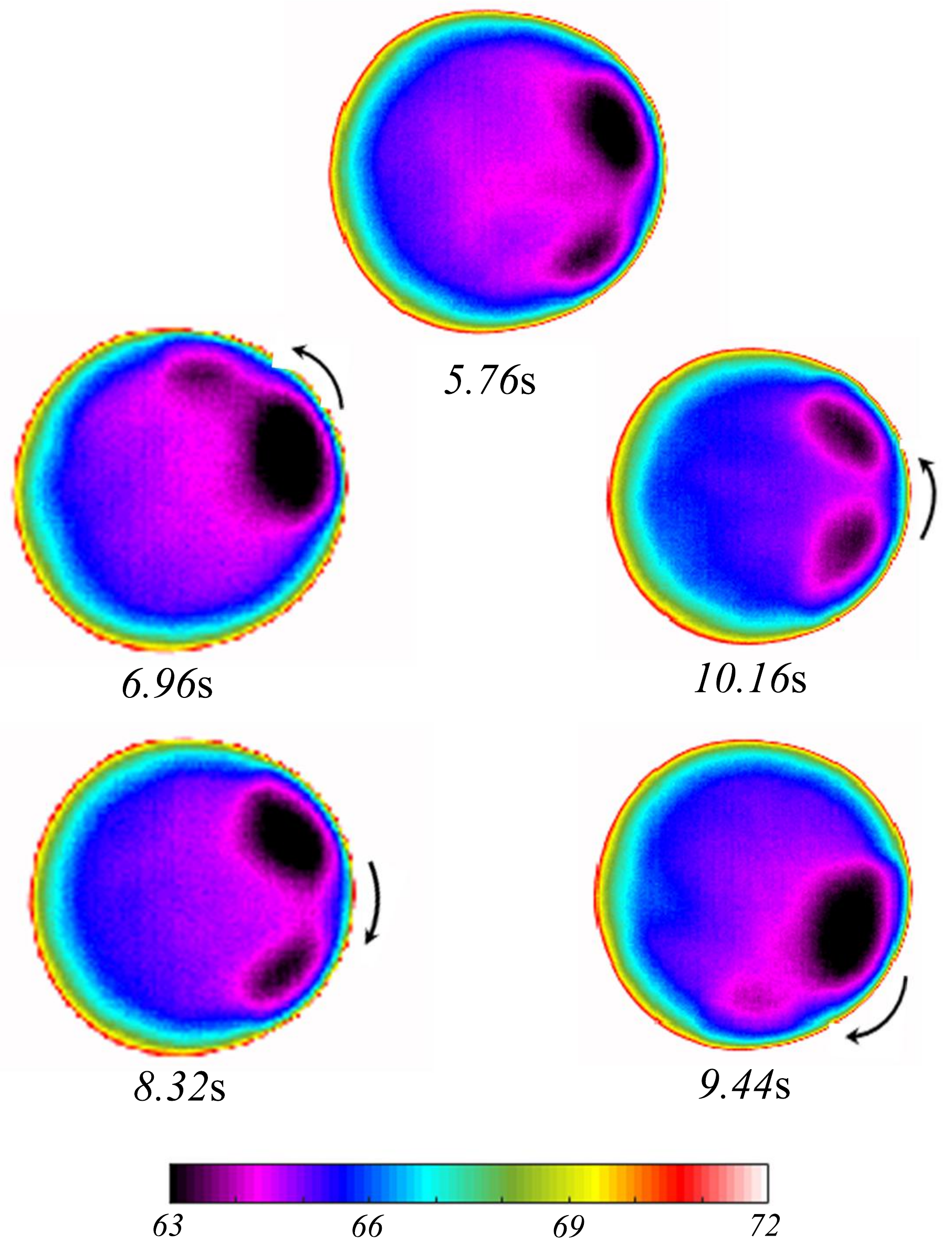}
\caption{Magnified views of the temperature contours at the early stage showing the oscillatory movement of the cold regions at $T_s=70^\circ$C.}
\label{fig:figS2}
\end{figure}

\begin{figure}
\centering
\includegraphics[width=0.95\textwidth]{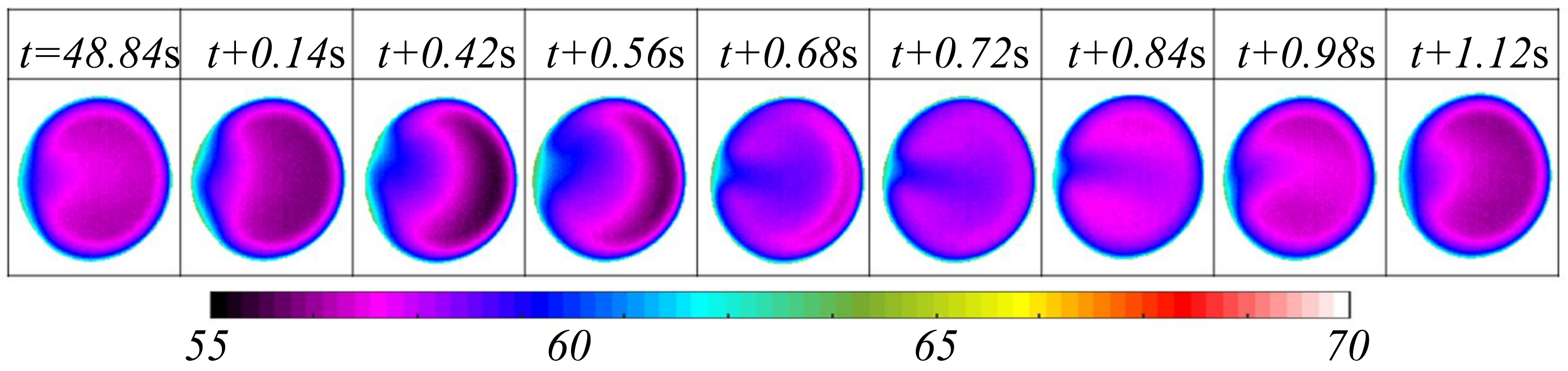}
\caption{A binary (E 20\%  + W 80\%) droplet exhibiting pulse generation near the receding contact angle of the droplet at $T_s=70^\circ$C and $\alpha=45^\circ$.   }
\label{fig:figS3}
\end{figure}

\begin{figure}
\centering
 \hspace{0.6cm} (a) \\
 \includegraphics[width=0.5\textwidth]{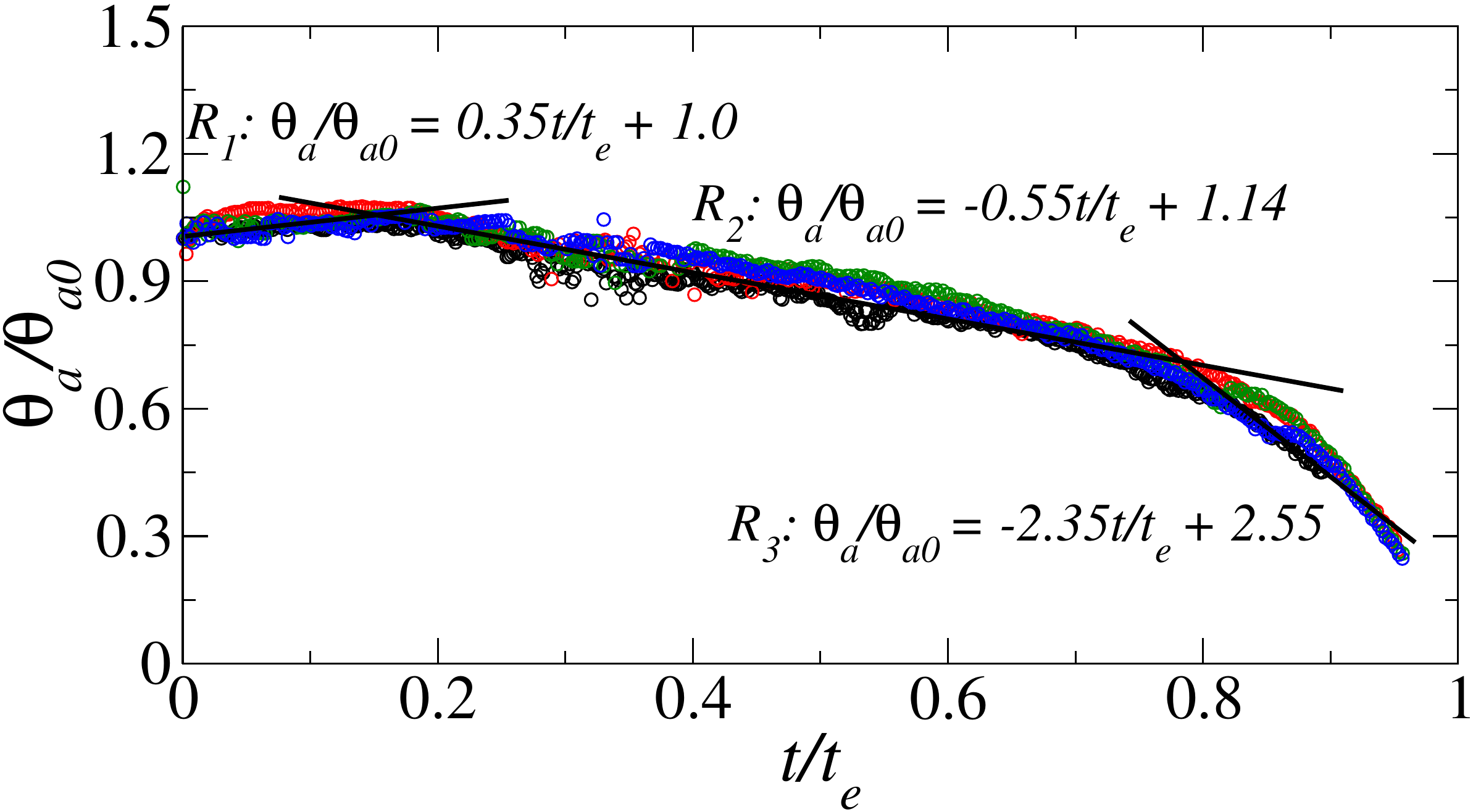} \hspace{2mm} \\
 \hspace{0.6cm} (b) \\
 \includegraphics[width=0.5\textwidth]{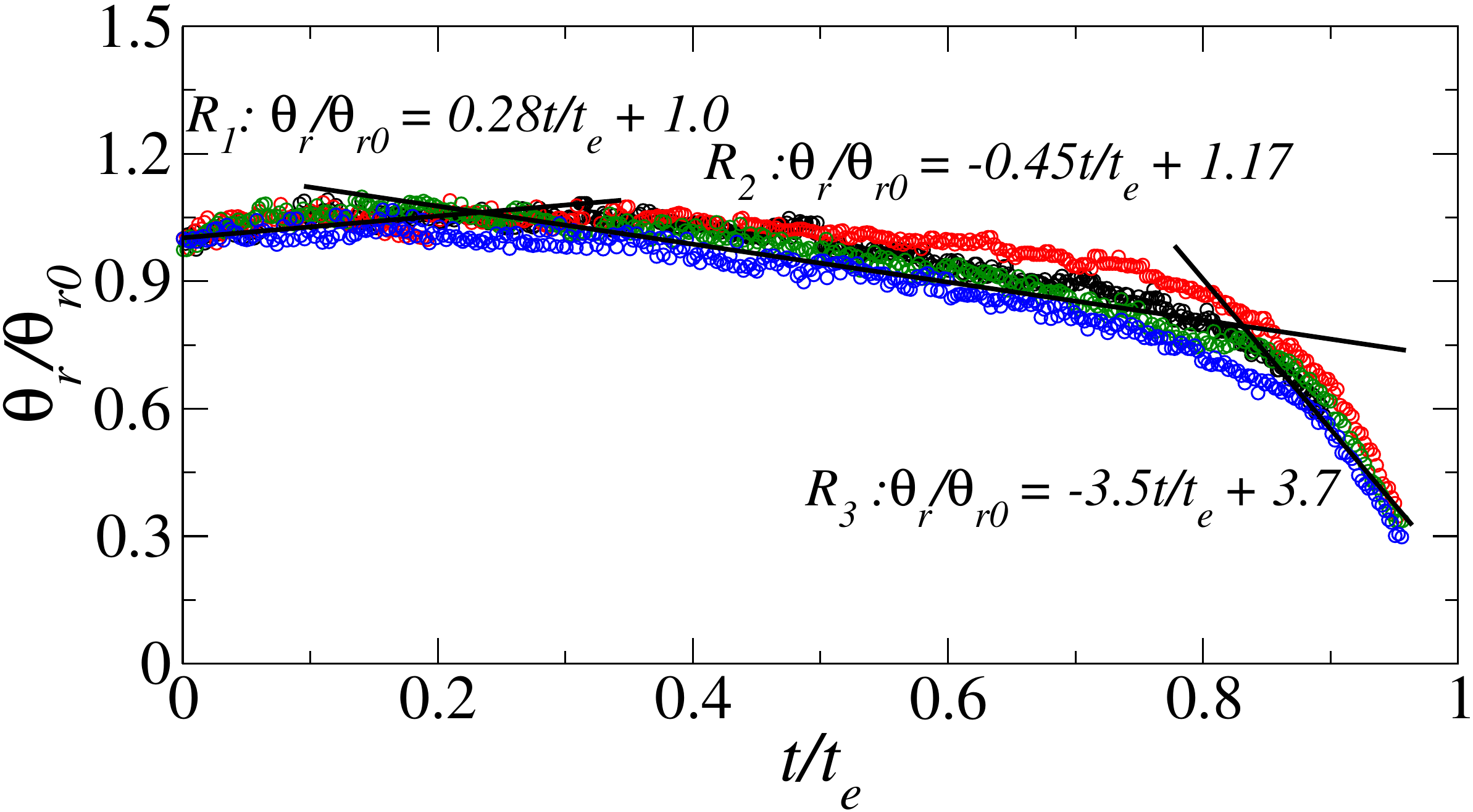}\\
\caption{Demonstration of three different regions of the (a) normalized advancing $(\theta_a/\theta{a0})$ and (b) normalized receding $(\theta_r/\theta{r0})$  contact angles of a (E 20\%  + W 80\%) droplet with the corresponding initial angles at $T_s =50^{\circ}$C and $\alpha = 40^{\circ}$. Here, $\theta_{a0}$ and $\theta_{r0}$ are the advancing and receding contact angles at $t=0$.}
\label{fig:figS4}
\end{figure}

\begin{table}
\caption{The values of the slope and intercept in different regimes associated with the (a) advancing and (b) receding contact angles at different substrate temperatures, as shown in Figs. 11(c) and (d), respectively.} \label{T1}
	\vspace{5mm}
 \hspace{3.0cm}  (a) \hspace{5.0cm} (b) \\
	\begin{minipage}{0.5\textwidth}
	\centering
	\resizebox{\textwidth}{!}{
		\begin{tabular}{|c|c|c|c|} \hline
		$T_s$ ($^{\circ}$C)  &    Slope  &  Intercept & Region  \\ \hline
\multirow{3}{*}{40} &    0.88   &  0.95      & R1      \\ \cline{2-4} 
                &   -0.8    &  1.34      & R2      \\ \cline{2-4}   	
		        &   -2.92   &  3.03      & R3      \\   \hline
\multirow{3}{*}{50} &  0.35     & 1          & R1      \\ \cline{2-4} 
		        &  -0.55    &  1.14       & R2      \\ \cline{2-4} 
		        & -2.35     &  2.55       & R3      \\ \hline
\multirow{3}{*}{60} & 0.45      & 1          & R1      \\ \cline{2-4} 
	            & -0.5      & 1.2        & R2       \\ \cline{2-4}  
	            & -3        &  3.25      & R3       \\ \hline 
\multirow{3}{*}{70} & 0.5       & 1          & R1       \\ \cline{2-4} 
				& -0.63  	& 1.21 		 & R2  		\\ \cline{2-4} 
				&  -2.5     &   2.7      & R3       \\ \hline
	\end{tabular}
	}
	\label{(a)}
	\end{minipage}
	\begin{minipage}{0.5\textwidth}
	\centering
	%\hspace{0.6cm}  (a) (b) \\
	\resizebox{\textwidth}{!}{
		\begin{tabular}{|c|c|c|c|}
		\hline
		$T_s$ ($^{\circ}$C) &    Slope  &  Intercept & Region  \\ \hline
		\multirow{3}{*}{40} & 0.93 &  0.98 & R1 \\ \cline{2-4} 
	&   -0.59                &  1.31 & R2  \\ \cline{2-4}   
	&   -3.56   &  3.72 & R3  \\  \hline
		 \multirow{3}{*}{50} &  0.28 & 1   &R1 \\ \cline{2-4} 
		 & -0.45               &  1.17 & R2  \\ \cline{2-4} 
		& -3.5   &  3.7 & R3  \\  \hline
	   \multirow{3}{*}{60} & 0.8 & 1  &R1 \\ \cline{2-4} 
	    & -0.5     &1.18    &R2       \\ \cline{2-4}  
	& -4.5      &  4.7 & R3 \\  \hline
		 \multirow{3}{*}{70} &0.5  &1  &R1\\ \cline{2-4} 
		& -0.5  & 1.28 &R2  \\ \cline{2-4} 
		&  -2.5         &   2.82 & R3 \\ \hline
	\end{tabular}
	}
	\label{(b)}\hfill
	\end{minipage}
\end{table}

\clearpage
\subsection{Evaluation of droplet volume and free surface area} \label{A2}
The volume and the free-surface area calculations are different from that used in case of a flat plate. The contour of droplet is asymmetric in case of an inclined plate. The droplet side contours at different times are obtained during drop evaporation as shown in Fig. 11. A MATLAB code is developed to calculate the droplet volume and the surface area with time from these contours. Fig. S5 shows a schematic of the droplet contour at a particular instant. The apex of the drop, point A, is identified where the droplet height attains the maxima. The mid-point of the wetted base, point O, is also obtained. The line joining point O with point A is along the drop axis OO$^\prime$. The entire drop is divided into elements of differential thickness dx along the length OA.

\begin{figure}
\centering
\includegraphics[width=0.8\textwidth]{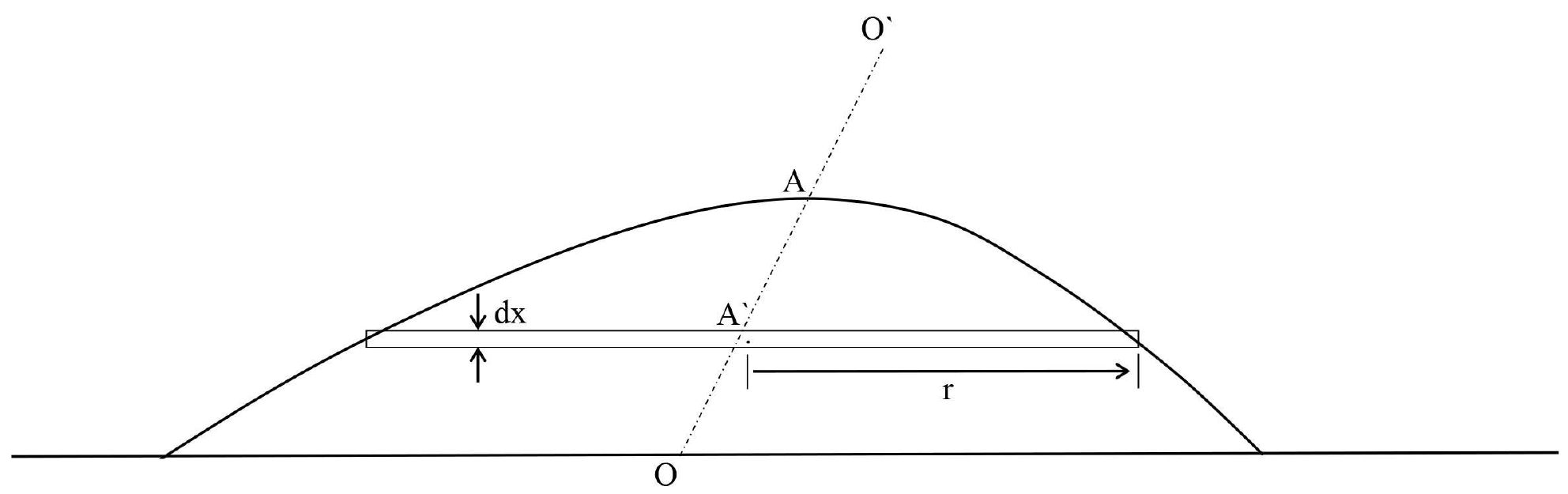}
\caption{Schematic representation of the volume and surface area calculations of a droplet on an inclined substrate.}
\label{fig:figS5}
\end{figure}

The value of $dx$ is very small and it is a fair assumption to consider this small element dx to be straight vertically such that we can use the volume and surface area expressions of a cylinder of radius 'r' and thickness 'dx'. 
\begin{equation}
dx = {OA \over n}
\end{equation}
where $n$ = number of divisions.
The volume of this small element is calculated as 
\begin{equation}
dV = {\pi r^2 dx }
\end{equation}
The surface area of the small element is calculated as
\begin{equation}
dA_s = {2\pi r dx }
\end{equation}
Starting from the base of the drop the volume of each element is calculated and is summed up to obtain the total volume of the drop V and the total surface area $A_s$.
\begin{equation}
V = {\sum_{i=1}^{\infty} dV_i}
\end{equation}
\begin{equation}
A_s = {\sum_{i=1}^{\infty} dA_{s_i}}
\end{equation}
For each successive iteration, the value dx is added along the axis OO$^\prime$ and the calculation is repeated. The number of divisions  $n$ is chosen based on a convergence study to make the calculation more robust.

\begin{figure}
\centering
\includegraphics[width=0.5\textwidth]{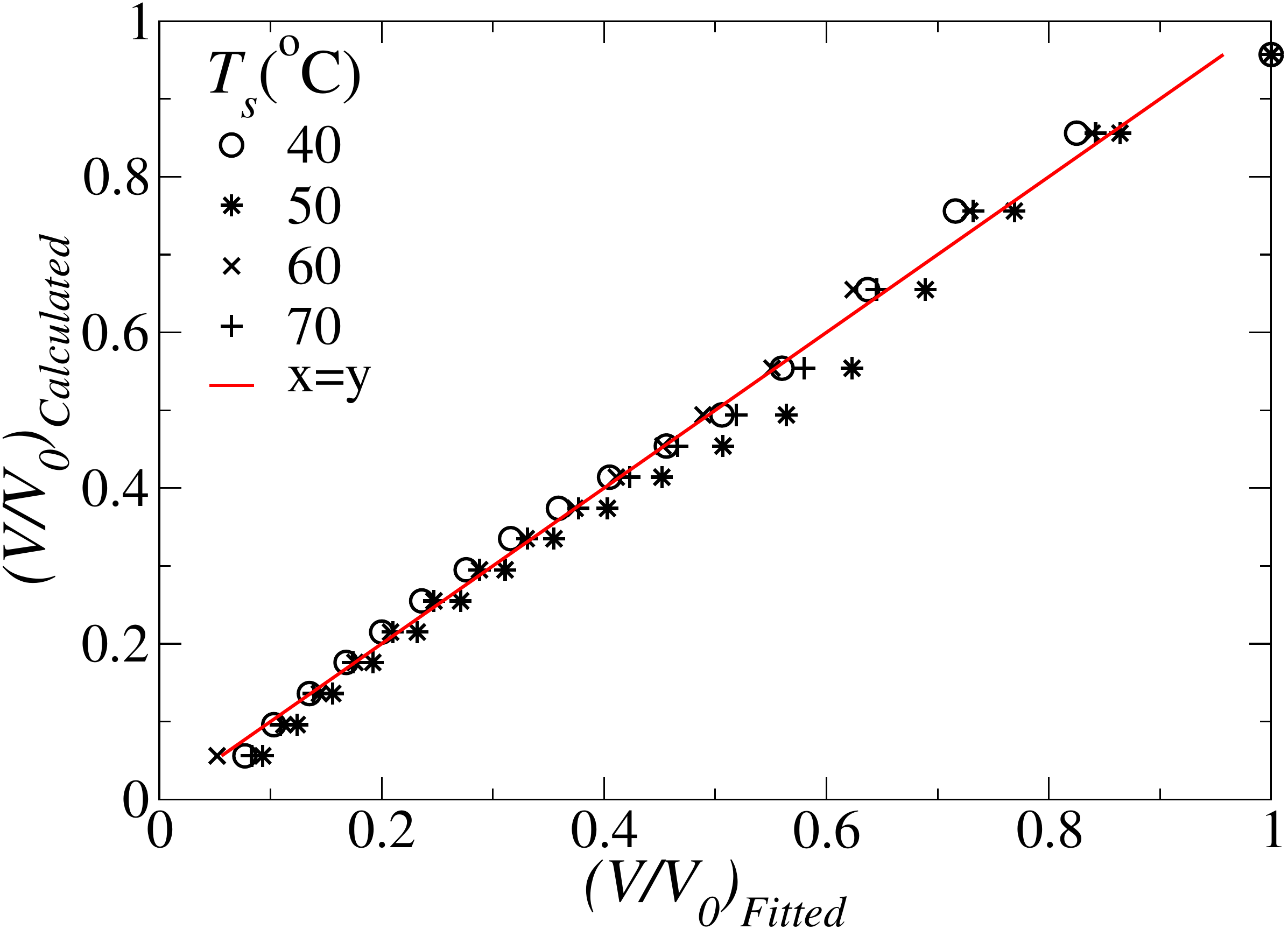}
\caption{Scatter plot of the fitted and experimentally calculated binary droplet volumes at different substrate temperatures.} 
\label{fig:figS6}
\end{figure}

\end{document}